\documentclass[12pt,preprint]{aastex}

\begin{document}

\shorttitle{Chandra Obs. of h Persei}
\shortauthors{Currie, T., et al.}

\title{The X-Ray Environment During the Epoch of Terrestrial Planet Formation: Chandra Observations of
h Persei}
\author{Thayne Currie\altaffilmark{1}, Nancy Remage Evans\altaffilmark{1}, Brad D. Spitzbart\altaffilmark{1},
Jonathan Irwin\altaffilmark{1}, Scott J. Wolk\altaffilmark{1}, Jesus Hernandez\altaffilmark{2, 4},
Scott J. Kenyon\altaffilmark{1}, and Jay M. Pasachoff\altaffilmark{3}}
\altaffiltext{1}{Harvard-Smithsonian Center for Astrophysics, 60 Garden St. Cambridge, MA 02140}
\altaffiltext{2}{Department of Astronomy, University of Michigan}
\altaffiltext{3}{Department of Astronomy, Williams College}
\altaffiltext{4}{Centro de Investigaciones de Astronom\'{\i}a, Apdo Postal
264, Merida 5101-A, Venezuela}
\email{tcurrie@cfa.harvard.edu}
\begin{abstract}
We describe Chandra/ACIS-I observations of the massive $\sim$ 13--14 Myr-old cluster, 
h Persei, part of the famous Double Cluster (h and $\chi$ Persei) in Perseus.  Combining 
the list of Chandra-detected sources with new optical/IR photometry and optical spectroscopy 
reveals $\sim$ 165 X-ray bright stars with V $\lesssim$ 23.  Roughly 142 have optical magnitudes 
and colors consistent with cluster membership.  The observed distribution of L$_{x}$ peaks 
at L$_{x}$ $\sim$ 10$^{30.3}$ ergs s$^{-1}$ and likely traces the bright edge of a far larger population 
of $\approx$ 0.4--2 M$_{\odot}$ X-ray active stars.
From a short list of X-ray active stars with IRAC 8 $\mu m$ excess from warm, terrestrial-zone dust, we derive a  
maximum X-ray flux incident on forming terrestrial planets.  Although there is no correlation 
between X-ray activity and IRAC excess, the fractional X-ray luminosity correlates with
optical colors and spectral type.  By comparing the distribution of L$_{x}$/L$_{\star}$ 
vs. spectral type and V-I in h Per with results for other 1--100 Myr-old clusters, we show that stars 
slightly more massive than the Sun ($\gtrsim$ 1.5 M$_{\odot}$) fall out of X-ray saturation by $\approx$ 10--15 Myr.
Changes in stellar structure for $\gtrsim$ 1.5 M$_{\odot}$ stars likely play an important role in this 
decline of X-ray emission.
\end{abstract}
\textit{Facility}: \facility{Chandra X-Ray Observatory, Kitt Peak National Observatory, 
MMT Observatory, Spitzer Space Telescope}
\section{Introduction}
Intermediate and low-mass main sequence stars like the Sun 
show evidence for chromospheric/coronal activity. 
Diagnostics of this activity -- Ca II H and K emission, X-ray emission, etc. -- 
are likely produced from a self-sustaining magnetic dynamo, which 
results from a combination of convective energy transport and 
differential rotation \citep{Pa55, Ba61, Ba03}.   
Once stars begin to contract onto the main sequence, the
X-ray luminosity (L$_{x}$) correlates with stellar rotation (v), 
with L$_{x}$ $\propto$ v$^{2}$ \citep{Pa81, Gu97}.  This evolution provides 
a quantifiable link between a major diagnostic of activity and the source 
of activity \citep[see also][]{Pi03}.  Therefore, studying the evolution 
of X-ray emission provides an insight into stellar structure and evolution 
and the influence of activity on the circumstellar environment.  

Young analogs of main sequence field stars have much higher levels 
of X-ray emission than main sequence stars.  X-ray surveys of young, 
$\lesssim$ 100 Myr-old  open clusters reveal that intermediate mass
($\sim$ 1-3 M$_{\odot}$) and low-mass stars are strong X-ray emitters.  
Typical  X-ray luminosities for young stars are $\sim$ 100-1,000 times 
larger than the characteristic solar X-ray luminosity \citep{Fe05, Pr05, Pr205}.  

X-ray observations of clusters (e.g. with Chandra) provide detailed 
constraints on the time history of chromospheric activity.
  As stars age and evolve onto the main 
sequence, their magnetic activity diminishes.  Thus, their X-ray luminosity 
declines \citep{Sk72, Mi85, Gu97, Fe99, Es00, Pr05}.  This decay is likely 
due to the rotational spin-down of the star and reduction of differential rotation 
at the radiative/convective zone interface \citep[e.g.][]{No84, Ba95}.
The youngest ($\sim$ 1 Myr old) stars less massive than $\sim$ 1-2 M$_{\odot}$ 
typically have L$_{x}$ $\sim$ 10$^{29}$--10$^{31}$ ergs s$^{-1}$ and fractional 
luminosities (L$_{x}$/L$_{\star}$, where L$_{\star}$ is the bolometric luminosity) of $\sim$ 10$^{-3}$--10$^{-4}$ \citep{Pr05}.  
Observations of older clusters like the Pleiades \citep{Mi99} suggest that the 
typical X-ray luminosities drop to 10$^{29}$ ergs s$^{-1}$ and fractional 
luminosities drop to $\sim$ 10$^{-4}$--10$^{-4.5}$ by 100 Myr.

Observations of young stars also show that a stellar mass
dependence of X-ray emission emerges sometime after $\approx$ 2 Myr.
In the COUP observations of the Orion Nebula Cluster,  young stars have a 
wide range of fractional luminosities which are not correlated with stellar 
mass \citep{Pr05, Pr205}.  X-ray activity from the magnetic dynamo is  
uncorrelated with rotation rate and is described as being "saturated", L$_{x}$/L$_{\star}$ $\sim$ 10$^{-3}$, 
or "supersaturated", L$_{x}$/L$_{\star}$ $>$ 10$^{-3}$ 
 \citep{Ra96, Gu04}.  In contrast, observations of older 
clusters like the Pleiades suggest clear relations between L$_{x}$/L$_{\star}$ 
and stellar mass \citep{Mi99}, where lower-mass stars have a higher fractional 
luminosity.  Thus, sometime between 1 Myr and 100 Myr, stellar rotation rates 
decrease sufficiently to bring the X-ray activity out of saturation 
\citep{Ps96, Gu04}.  

As stars evolve to the main sequence, their X-ray emission may drive important processes in the circumstellar disk 
material from which planets form.  X-ray active stars with ages of $\sim$ 1-10 Myr 
 are surrounded by primordial circumstellar disks comprised of 
gas and small dust grains \citep{Kh95}.  X-ray irradiation may provide an ionization 
source for primordial disks that powers disk viscosity to sustain disk accretion when MRI turbulence is 
otherwise ineffective \citep{Gl97}.  Flash heating from X-rays in powerful solar flares may account 
for the formation of chondrules in the early solar nebula \citep{Sh96}.  X-ray 
heating of solids during flaring events, perhaps even during the early debris 
disk stage, may also drive chemical reactions from spallation which produces the short-lived 
isotopes observed in some meteorites \citep{Fei02}.  

X-ray irradiation also has important consequences for the evolution of newly-formed planets.  Large 
X-ray fluxes can ablate the hydrogen-rich atmospheres of short-period giant planets \citep{La03} 
and remove the atmosphere entirely for sufficiently high L$_{x}$ (see also \citealt{Ba05}).  
For an X-ray luminosity function scaled to the Pleiades, nearly all ($\sim$ 85\%) Neptune-mass gaseous 
planets at 0.02 AU from a solar-type star may be eroded to Super-Earths \citep{Pe08}.  
G-type stars with high X-ray luminosities can evaporate the atmospheres of a wide range of planet 
masses ($\sim$ 0.1-10 M$_{\oplus}$) within $\sim$ 0.1 AU.  In the absence of evaporation, X-ray 
irradiation may also drive a variety of photochemical reactions in the atmospheres of 
terrestrial planets \citep[e.g.][]{La07}.  

Understanding the evolution of stellar activity and the X-ray environment of circumstellar
disks and planets requires observations of stars in clusters with ages of $\approx$ 10--30 Myr.
Recent X-ray surveys have concentrated on $\lesssim$ 5 Myr-old clusters such as Orion, NGC 2264, and 
IC 348 and $\gtrsim$ 50--100 Myr old clusters such as $\alpha$ Persei and the Pleiades 
\citep{Pr96, Mi99, Pr02, Pr05, Dahm07}.  The large gap in ages between the youngest and the oldest clusters
makes it difficult to constrain the evolution of the X-ray luminosity function and the evolution
of stellar activity.  X-ray data for 10--30 Myr-old clusters also link constraints on the 
X-ray environment during accretion and grain growth in the primordial disk phase with those 
for $\gtrsim$ 100 Myr-old stars where planet formation should be complete in the terrestrial zone.  

In this paper, we report analysis of Chandra observations of 13--14 Myr old 
h Persei, part of the famous Double Cluster h and $\chi$ Persei.  
Recent Spitzer observations of h Persei \citep{Cu07a, Cu08a,Cu08b} reveal that the 
cluster harbors a substantial population of stars with warm dust emission 
consistent with debris from active terrestrial planet formation \citep{Cu07b, Cu08b}.  
At $\sim$ 13-14 Myr \citep{Sl02, Me93, Cu08b},
h Persei probes ages during which terrestrial planets
are potentially in the final stages of formation \citep{Kb06, Ws93}
and when nebular gas has recently dissipated (\citealt{Cu07c}, see also \citealt{Th08} and references therein), leaving
newly-formed gas giants unshielded from high energy X-ray photons.

Our main goals for this paper are twofold.  First, we investigate 
the connection between X-ray activity and warm debris emission.  For 
X-ray bright stars with evidence for warm dust, we 
estimate an X-ray flux in the terrestrial zone that provides 
input for planetary atmosphere evaporation models.  Second, we provide 
constraints on the time-evolution of stellar X-ray activity linking studies 
of younger $\sim$ 1--2 Myr-old clusters like Orion and older clusters like 
the Pleiades.  Chandra sensitivity limits likely preclude sampling the 
full distribution of X-ray active stars in h Persei.  However, we can 
analyze the upper envelope of the distribution of L$_{x}$ and L$_{x}$/L$_{\star}$ as a function of 
stellar properties.  From comparing the bright limit of the L$_{x}$/L$_{\star}$ distribution from h Per 
with the distribution from younger and older clusters, we look for evidence of evolution in 
stellar activity.  A detailed investigation of the X-ray spectra for sources with the highest X-ray 
counts will be included in future work (N. Evans et al., in prep.)

\section{Chandra Observations and Ancillary Data}
\subsection{Observations, Image Processing, and Source Extraction}
Chandra observations of h Persei were taken with a 41.1 ksec exposure 
on Dec. 2, 2004 (Obs. ID 5407; Sequence Number 200341) with 
the ACIS detector \citep[chips 0, 1, 2, 3, 6, and 7;][]{We02}.  
The data were obtained in dithered, timed mode, with a 
frametime of 3.2 seconds.  On-board event rejection and event telemetry was in VFAINT mode.
The field was centered on $\alpha_{2000}$=2$^{h}$19$^{m}$00$^{s}$, $\delta_{2000}$=57$^{o}$07'12",  
close to the center of h Persei from \citet{Bk05} 
($\alpha_{2000}$=2$^{h}$18$^{m}$56.4$^{s}$, $\delta_{2000}$=57$^{o}$08'25") and observed 
at a roll angle of 229$^{o}$.  The data were not registered to an astrometric 
reference frame (e.g. 2MASS).  The ACIS-I field covers a 17'x17' area.

\begin{figure}
\centering
\plotone{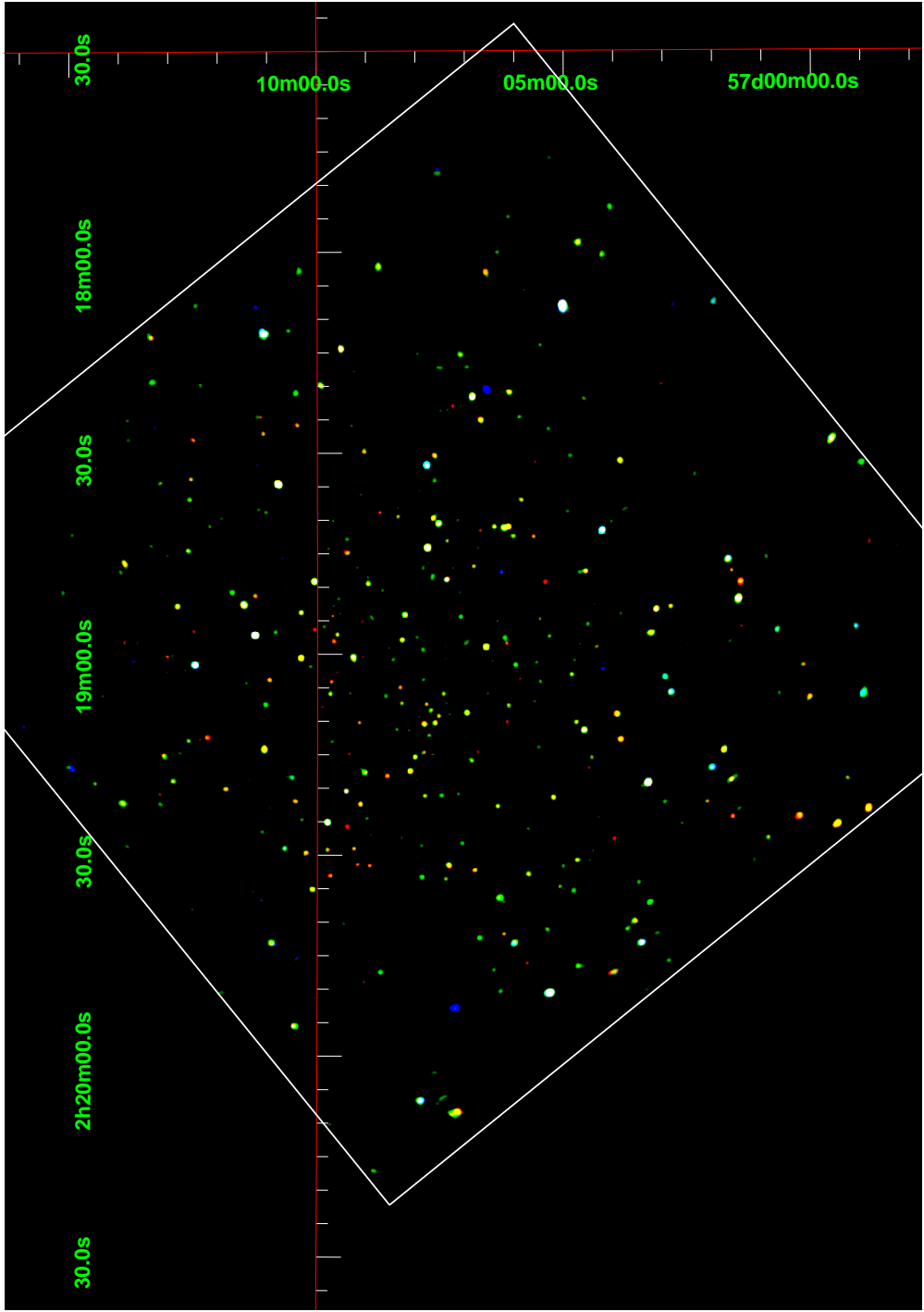}
\caption{False-color image of the Chandra field binned by 4 pixels and convolved with a 3-pixel gaussian.  
The center of h Persei is at $\alpha_{2000}$ $\sim$ 2$^{h}$18$^{m}$56.4$^{s}$, $\delta_{2000}$ $\sim$ 57$^{o}$8'25" 
according to \citet{Bk05}.}
\label{image}
\end{figure}

Figure \ref{image} shows the reduced, smoothed image, binned by 4 pixels and then 
convolved with a 3 pixel gaussian to balance the prominence of the sources in the 
center of the field with those on the edges.  On the original
data (which is the basis for all further analysis) the 
 sources in the center of the field are smaller
 than sources on the edge where the point spread function (psf)
is larger.  

We process the images with the standard ANCHORS (An Archive of Chandra
Observations  of Regions of Star  Formation)  pipeline (\url{http://cxc.harvard.edu/ANCHORS/}; 
\citealt{Sp08}).  Based on the CIAO (Chandra Interactive Analysis of Observations) routines, 
this reduction system is designed to process Chandra images containing
many point sources and diffuse emission.  
The CXC pipeline level 2 data products (DS version 7.6) required no 
reprocessing; we only applied an energy filter at 0.3--8.0 keV to eliminate the high-energy background. 

To select X-ray sources, we investigated several 
source detection algorithms available in CIAO.
We used a recursive blocking scheme with WAVDETECT \citep[Mexican hat wavelet 
detection]{Free02} with the significance set to yield about 1 false detection per field.  
A 15'x15' region centered on the aimpoint was extracted at full resolution.  
WAVDETECT then identified sources in this region using wavelet scales of 
2, 4, 8, 16, and 32.  
We merged the three lists, removing any duplicates.  

This method yields 330 point source detections.  
Source positions were then adjusted using a centroiding routine.
We extracted counts using an elliptical extraction region around each source 
based on point-spread function fitting as a function of chip position 
at 95\% encircled energy.  Background levels were computed using 
an annular ellipse with the same shape as the source extraction.  The outer 
ellipse boundry has major and minor axes six times the size of the source; the 
inner boundry has axes three times the size of the source.  
\subsection{Deriving the X-Ray Flux and Temperature}
We derived the X-ray flux and temperature for each Chandra source 
using a one temperature APEC model with absorption using the CIAO fitting and modeling 
package Sherpa, v. 3.4\footnote{http://cxc.harvard.edu/sherpa3.4/index.html/}.  
Unless X-ray sources are very strong (100 counts or more),
it is useful to fix several parameters in the fits to
prevent unphysical values.  The fixed parameters are the hydrogen column 
density, n$_{H}$, the initial X-ray temperature, and the abundance parameter.  

We use the well-constrained reddening of h Persei to set 
n$_{H}$.  Based on $\sim$ 100 high-mass stars, \citet{Sl02} derive E(B-V) $\sim$ 0.56 $\pm$ 0.01, 
a value within 10\% of estimates from similar work \citep[e.g.][]{Ke01, Bk05} and identical to that derived 
for $\chi$ Persei.  From optical spectroscopy of $\sim$ 6,000 stars, \citet{Cu08b} and \citet{Cu08c}  
derive E(B-V) $\sim$ 0.56 for both h and $\chi$ Persei.  \citet{Cu08b} identifies 
a full-width half-maximum of $\sim$ 0.1 in the distribution of E(B-V).  
Very few cluster stars have E(B-V) $\gtrsim$ 0.65 or $\lesssim$ 0.45.  
Because E(B-V) is proportional to n$_{H}$, adopting a single 
n$_{H}$ $\approx$ 3 $\times$ 10$^{21}$ cm$^{-2}$ 
for E(B-V) $\sim$ 0.56 introduces a $\approx$ 10\% uncertainty in n$_{H}$ and a $\approx$ 0.05 dex 
uncertainty in log(L$_{X}$) for each source.  
Additionally, there is a $\approx$ 10\% systematic uncertainty as we chose 
a constant of proportionality for the E(B-V) to N$_{H}$ relation which is intermediate 
between the values of \citet{Vuong2003} and \citet{Ryter1996}.

We adopted an initial temperature kT of 1.5 keV as a compromise 
between values typical of stars in younger clusters 
(e.g. M17, \citealt{Broos07}) and stars in older clusters (e.g. the Pleiades, 
\citealt{Dlg02}).  
An abundance parameter of 0.3 solar is routinely found in fits of X-ray 
spectra and assumed constant in our calculations \citep{Fei02}.  The n$_{H}$ and 
abundance parameters remain fixed throughout the fitting, 
while kT is determined from fitting.  
The remaining free parameter is the normalization from which Sherpa derives 
both the absorbed and unabsorbed photon fluxes.  

We performed a two-step fitting procedure.  First, the unbinned data were fitted 
using C-statistics \citep{Cash79} and Powell optimization (identifies local fit statistic minimum 
nearest to initial guess.  For sources with more than 30 counts, the data were
then grouped for 8 counts/bin and refit using the result from the unbinned fit as an 
initial guess for the normalization and temperature.  The binned fitting used $\chi$-dvar statistics 
and Levenberg-Marquardt optimization.  The details of these optimization methods are described 
in the online Sherpa documentation\footnote{http://cxc.harvard.edu/sherpa/methods/methods.html}.

Table \ref{Chandradet} lists data for Chandra-detected sources.  Column 1 is the source ID, columns 2 
and 3 list J2000 positions, columns 4 and 5 list the raw and net counts in 
the 95\% encircled energy radius.
The hardness ratios (columns 6--8) describe the energy of 
X-ray photons.  Following \citet{Get05}, we use three hardness ratios.  
HR1 compares the full range of CXO sensitivity: 0.5--2.0 keV vs. 2.0--8.0 keV.  
HR2 highlights differences in the softer region (0.5--1.7 keV vs. 1.7--2.8 keV), while 
HR3 highlights differences in the harder regions (1.7--2.8 keV vs. 2.8--8.0 keV).
Columns 10 and 11 list the X-ray temperature and uncertainty in temperature.
Although a detailed analysis of the X-ray spectra is the subject of future work (N. Evans et al., in prep.), 
the median X-ray temperature for the entire sample is $\sim$ 1.6 keV, which is reasonable for stars with ages 
intermediate between $\sim$ 1 Myr and 100 Myr as described in \S 2.2.
 Columns 12 and 13 list the derived unabsorbed and absorbed X-ray fluxes.  The median values for the unabsorbed and 
absorbed fluxes are 10$^{-14.6}$ ergs cm$^{-2}$ s$^{-1}$ and 10$^{-14.8}$ ergs cm$^{-2}$ s$^{-1}$.   
The reduced $\chi^{2}$ of the Sherpa fits are 
listed in column 14.
\subsection{Optical/IR Ancillary Data}
There are several possible types of X-ray active sources 
on the Chandra field, including chromospherically active pre-main 
sequence stars, active galactic nuclei (AGN), and foreground/background red giant stars. 
Without useful proper motion data for this distant cluster, we rely on 
optical/infrared colors and deep Chandra number counts \citep[e.g.][]{Bau04}
to identify likely cluster members and to make plausibility arguments for sources 
without optical IDs.  For populous clusters like h and $\chi$ Persei, we 
expect a well-defined locus in an optical color-magnitude diagram \citep[e.g.][]{Ly06}.
The optical luminosities for young (e.g 10--50 Myr)
intermediate/late spectral type pre-main sequence stars
(e.g. G0-M5) are up to $\approx$ 5 times greater than their main 
sequence field star counterparts \citep[e.g.][]{Ba98}.  Thus,
 field stars have far bluer colors at a given magnitude.  
Background AGN have very blue optical colors compared to 
pre-main sequence stars and have 
very faint optical magnitudes.  Foreground red giant stars 
have far redder colors than cluster stars at a given magnitude.
The differences in V/V-I positions for cluster stars and other X-ray 
active sources means that this color-magnitude diagram clearly
identifies the positions of X-ray active cluster members in moderately massive 
clusters such as NGC 2547 \citep[][]{Ly06}.  Because h Persei is
$\gtrsim$ 5--10 times more massive than NGC 2547, the cluster locus should be 
even more easily identifiable.

To identify the nature of Chandra-detected sources, 
we merged the Chandra catalog with deep VI optical photometry and 
optical spectra of h Persei\footnote{The optical photometry and 
spectroscopy presented in this section will be discussed in more detail
 in an upcoming paper \citep{Cu08c}.}.
Optical photometry of h Persei was taken with the Mosaic Imager at the 
4-meter Mayall telescope at the Kitt Peak National Observatory on October 13-16 and 
27-30, 2006, as a part of the MONITOR project \citep{Ai07}.
  Exposures in V and I band were taken using 75 second 
integrations with a 36'x36' field of view centered on the  
cluster.  The data were reduced using the 
pipeline for the INT wide-field survey \citep{Ir01, Ir07}, correcting for the effects of 
fringing, cross talk, bias, and atmospheric extinction.
  Photometry was performed as in \citet{Ir07}; 
instrumental magnitudes were transformed into Johnson-Cousins system magnitudes.
  The catalog contains $\approx$ 42,000 sources detected
in at least one band and are complete at the 5$\sigma$ level to V $\sim$ 23 and I $\sim$ 20.

For source matching, we use the optical photometry.
For 14 Myr-old cluster stars at the distance and reddening of h Per \citep[E(B-V) $\sim$ 0.56; d $\sim$ 2.3--2.4 kpc][]{Sl02, Cu08c}, 
stars with J-band magnitudes brighter than the 2MASS completeness limit (J $\sim$ 15.7) have 
V-I $\lesssim$ 1.75, V-J $\lesssim$ 3.3, and V $\lesssim$ 19 \citep{Kh95, Ba98, Si00}.
Cluster stars with V $\sim$ 23 (the 5$\sigma$ limit for our optical data) have V-I $\sim$ 3, 
 V-J $\sim$ 4.5, and J $\lesssim$ 18.5 \citep[cf.][]{Kh95}.  
Thus, the optical data detect fainter, lower-mass cluster stars than the 
2MASS data.

To provide more constraints on the properties of optically-detected Chandra sources, 
we include optical spectra of h Per sources.
Optical spectroscopy of $\approx$ 5,000 h Persei sources were taken during Fall 2006 and 
Fall 2007 with Hectospec \citep{Fa05} on the 6.5m MMT telescope.    
We took three 10-15 minute exposures using the 270 g mm$^{-1}$ grating.  This 
configuration yields spectra at 4000-9000 \AA\ with 3 \AA\ resolution.  
The data were processed using standard Hectospec reduction pipelines.  The 
reduced spectra typically had signal-to-noise $\gtrsim$ 20-40.  
Finally, we include $\sim$ 1000 archival spectra from the FAST and Hydra spectrographs on the 
1.5 m Tillinghast and 3.5 m WIYN telescopes, respectively.  \citet{Cu07b} and \citet{Cu08b} describe these 
data in detail.  The FAST and Hydra spectra consist mostly of bright, likely early-type stars in h Persei.

To derive spectral types from these spectra, we employed 
the semi-automatic quantitative spectral-typing code \textbf{SPTCLASS} from 
\citet[][see www.astro.lsa.umich.edu/hernandj/SPTclass/sptclass.html for more 
information]{He04}, which is useful for classifying stars regardless of luminosity 
class and surface gravity.  \textbf{SPTCLASS} calculates the spectral types of 
stars using spectral indices which compare the line flux of each spectral feature 
to continuum levels.  The relationships between spectral indices and spectral types 
are derived from spectroscopic standards observed with Hectospec.  Thus, \textbf{SPTCLASS} is 
particularly well suited to spectral type h Persei stars.  The errors in spectral types 
are typically $\approx$ 2 subclasses.

To identify the presence or absence of warm, circumstellar 
dust around X-ray active stars, we match ACIS sources to the 2MASS/IRAC catalog from \citet{Cu07a}.
\citet{Cu07a} describe the data reduction, photometry, and source matching between 2MASS and IRAC 
 in detail.  The catalog consists of $\sim$ 31,000 point 
sources and covers $\approx$ 0.75 square degrees on the sky, encompassing 
both h Persei and $\chi$ Persei.  The completeness limit for the catalog corresponds to the 
2MASS completeness limit (J $\approx$ 15.5); most stars brighter than J $\sim$ 15-15.5 have 
IRAC counterparts in at least one channel (3.6 $\mu m$, 4.5 $\mu m$, 5.8 $\mu m$, or 8 $\mu m$).

Of the 330 X-ray sources, 165 have optical photometric counterparts.
  The typical positional offsets are very small (mean offset is $\sim$ 0.60", 
median $\sim$ 0.47", and $\sigma$ $\sim$ 0.43") and comparable to the 
astrometric accuracy of Chandra.
Of the 165 X-ray/optically-detected sources, 101 have  
 spectral types;  123 have also 2MASS/IRAC photometry.
We derive the X-ray luminosity for optically-detected Chandra 
sources assuming a distance of 2.4 kpc.

The catalog of optically-detected Chandra sources along with their net X-ray counts 
and absorption-corrected X-ray luminosities\footnote{All 
X-ray luminosities discussed in the following sections are \textit{absorption-corrected} luminosities}, 
optical/IR photometry, and spectral types is listed in Table \ref{Chandralist}\footnote{The full versions of 
the tables are available in the electronic edition of this paper}.
\subsection{Nature of Chandra Sources Lacking Optical Detections}
By comparing the optical survey limits to colors and magnitudes for 
14 Myr-old pre-main sequence stars and other X-ray active sources 
we can constrain the properties of Chandra sources lacking optical 
detections.  The two main possibilities for these sources are 
fainter, low-mass cluster stars and background AGN.  
Simple arguments show that both fainter, low-mass cluster stars 
and AGN likely comprise the Chandra-detected population lacking optical detections.  

For the Geneva/Baraffe isochrone (see \S 3.1), 
 stars with V $\sim$ 22 have masses of $\sim$ 0.6 M$_{\odot}$ (Figure \ref{colorxray}).  
While model uncertainties plague our understanding of the optical colors/magnitudes 
of lower-mass stars, these stars should be cooler and fainter than 0.6 M$_{\odot}$ 
stars of the same age.  \citet{Ba98} predict that h Per stars with M$_{\star}$ $\sim$ 0.25 M$_{\odot}$ 
have V $\sim$ 24, two magnitudes fainter than 0.6 M$_{\odot}$ stars  
and undetectable with our optical data.  If these stars have high fractional 
x-ray luminosities ($\ge$ 10$^{-3}$, Chandra is likely sensitive enough 
to detect them.  Because the 5$\sigma$ limit 
for the optical data is V $\sim$ 23, stars of slightly greater mass ($\lesssim$ 0.5 M$_{\odot}$) 
may also lack optical detections but have Chandra detections.
 
Assuming a Miller-Scalo Initial Mass Function \citep{Ms79}, h Per stars with masses between 0.25 M$_{\odot}$ 
and 0.6 M$_{\odot}$ are 1.3 times as numerous as cluster stars with masses $\gtrsim$ 0.6 M$_{\odot}$.  
The number of optically-detected Chandra sources is equal to the number lacking optical detections.  
Therefore, lower-mass cluster stars plausibly comprise a significant 
fraction of the Chandra-detected population lacking optical detections.

Using the number counts of X-ray sources as a function of 
limiting flux from \citet{Bau04},  
 Chandra sources lacking optical detections likely include 
a significant population of optically faint, background AGN.  The limiting (unabsorbed) flux for our 
survey is log(F$_{x}$, ergs s$^{-1}$ cm$^{-2}$) $\approx$ -14.7.
\citet{Bau04} derive the number of
X-ray sources per sq deg with fluxes greater than
log (F$_{x}$ ergs s$^{-1}$ cm$^{-2}$) = -14.7 to be roughly 500 soft X-ray sources 
and 1000 hard X-ray sources.  Because the ACIS-I coverage is $\approx$ 0.08 square degrees, 
our h Per field should have $\sim$ 40 (80) soft (hard) AGN.  Most AGN with 
log F$_{x}$ $\sim$ $-14.7$ have faint optical magnitudes \citep[R $\gtrsim$ 22--23;][]{Bau04} 
and are thus fainter than the 5 $\sigma$ limit of our optical survey.

Other properties of sources with and without optical 
Several X-ray sources on the field have 2MASS detections but no detectable optical counterparts.  
These are typically very faint and do not define a locus in near-IR color-magnitude diagrams 
consistent with cluster membership.   These sources are then more likely to be AGN.

In summary, the Chandra sources lacking optical counterparts are likely to be a mix of 
low-mass cluster stars and background AGN.  Deeper optical data will likely 
identify the origin of some currently unmatched sources.  Deeper X-ray data will yield 
enough counts to fit a spectrum to many sources; analyzing their X-ray spectra 
can distinguish between pre-main sequence stars and AGN.
\section{Analysis}
\subsection{Optical Properties of Chandra-detected Sources}
\begin{figure}
\plotone{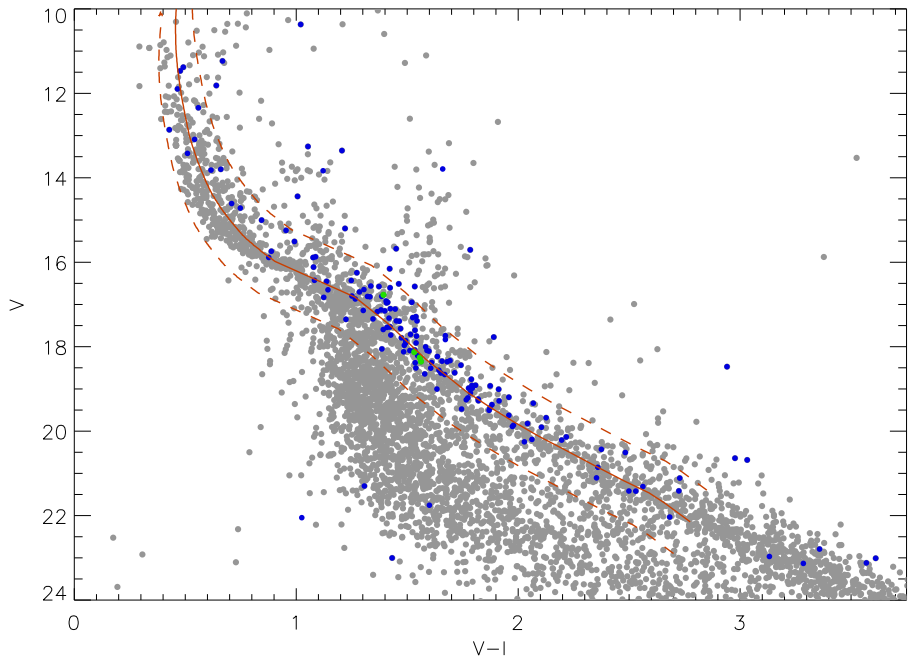}
\caption{V/V-I color-magnitude diagram with h Persei sources within 10' of the cluster center (grey dots).  
Blue dots correspond to Chandra-detected sources with K$_{s}$-[8] $\le$ 0.5 and light-green dots 
correspond to Chandra-detected sources with K$_{s}$-[8] $\ge$ 0.5.  The solid line denotes the best-fit (14 Myr) isochrone 
using the Geneva/Baraffe models with 0.75 magnitude upper and lower limits.  The lower mass limit 
for the isochrone (V=22, V-I=2.8) corresponds to 0.6 M$_{\odot}$.  Most Chandra sources 
with optical counterparts follow the predicted isochrone.}
\label{colorxray}
\end{figure}

Most of the 165 optically-detected X-ray sources define a clear sequence in the V/V-I color-magnitude 
diagram consistent with cluster membership.  Figure \ref{colorxray} shows their positions (blue and green dots)
with respect to all sources (grey dots) detected within 10' of the h Persei center 
($\alpha_{2000}$ $\sim$ 2$^{h}$19$^{m}$0$^{s}$, $\delta_{2000}$ $\sim$ 57$^{o}$8'35"; see \citealt{Bk05}).
The cluster sources define a sequence from V,V-I $\sim$ 10,0 to V,V-I $\sim$ 24, 3.6, which 
is clearly separable from the background field star population (V $\gtrsim$ 15 + 3(V-I) for V-I $\gtrsim$ 1). 
Foreground stars are less numerous than either the cluster stars or the background.  The brightest 
X-ray active star (V $\sim$ 8) is located beyond the plot limits.

To compare the magnitudes and colors of these stars with theoretical predictions, 
we overplot a best-fit isochrone (14 Myr), covering a range in masses from 15 M$_{\odot}$ 
to 0.6 M$_{\odot}$.  We construct our isochrone for intermediate/low-mass stars
 ($\lesssim$ 1.4 M$_{\odot}$) 
 by interpolating between the 12.6 Myr and 15.8 Myr isochrones from \citet{Ba98}.  
  For more massive stars, we use the Geneva isochrones \citep{Sch92, Me93}.  
On Figure \ref{colorxray}, the 
\citeauthor{Ba98} and Geneva predictions meet at V $\sim$ 17, V-I $\sim$ 1.2.  
We adjust the isochrone to the distance and reddening of h Persei, 
assuming a distance modulus of 11.85 and E(B-V) $\sim$ 0.56 \citep{Sl02}.
For the rest of the paper, we label the isochrone constructed from 
tracks for high-mass stars from the Geneva group and intermediate/low-mass stars 
from Baraffe as the 'Geneva/Baraffe' isochrone.

The agreement between the predicted 14 Myr isochrone and the color-magnitude diagram 
for all h Per stars is exceptional.
The isochrone accurately reproduces the cluster locus from V, V-I = 10, 0 
to V, V-I = 22.5, 2.8.  At 14 Myr, this range corresponds to 
stars with masses $\gtrsim$ 0.6 M$_{\odot}$.
  Isochrones for 12.6 Myr and 15.8 Myr (not shown) also show fair agreement, though 
slightly younger (11.2 Myr) and older (17.8 Myr) isochrones provide a much poorer 
fit\footnote{A more detailed analysis of the success/failure of isochrones of different ages 
as well as isochrones from different groups will be included in an upcoming 
 paper \citep{Cu08c}}. 

The vast majority (142/165)
of Chandra-detected sources, especially those fainter than V=16, track this sequence extremely well.  
Many X-ray sources follow the isochrone from V $\approx$ 10 to V $\approx$ 21.  
A smaller number of much fainter X-ray sources (V $\sim$ 22--23, 
V-I $\sim$ 2.75-4) also track the locus of cluster stars.  In addition to 
stars in the cluster, 12 foreground stars are X-ray active.

\begin{figure}
\plotone{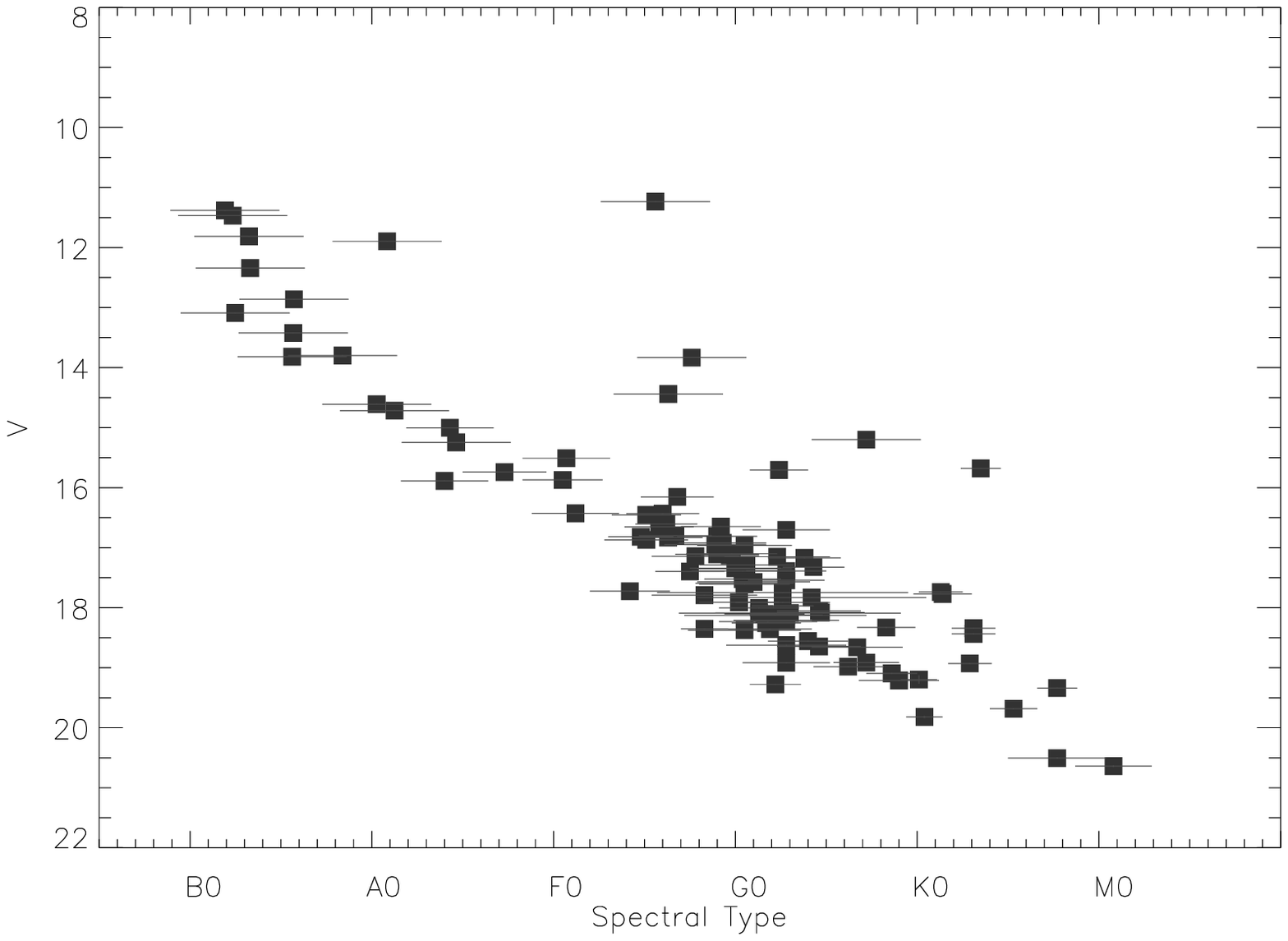}
\plotone{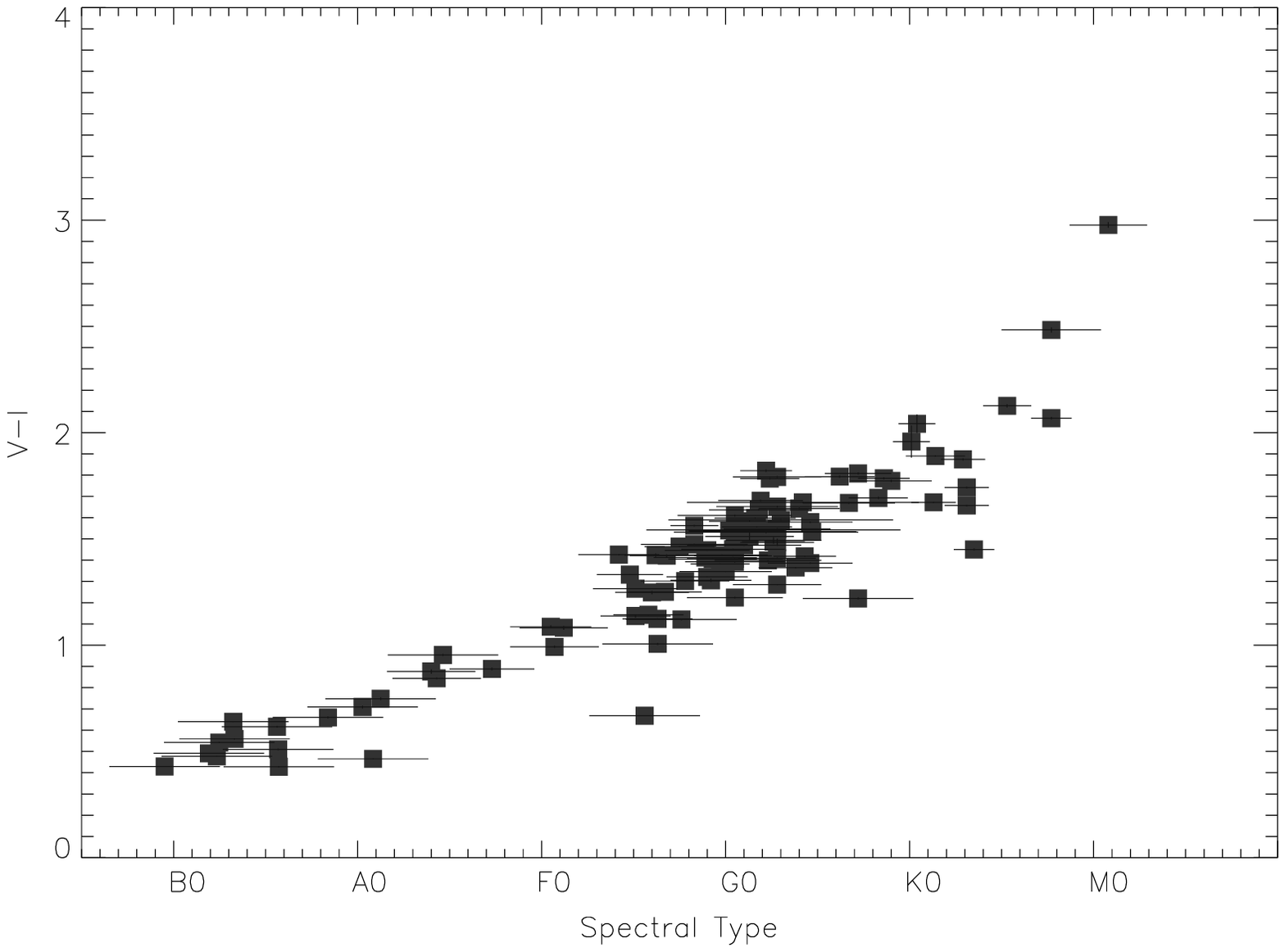}
\caption{V vs. spectral type (top panel) and V-I vs. spectral type (bottom panel) for Chandra-detected sources.  The sequence 
of X-ray active stars clearly traces a locus in both diagrams.  We overplot error bars for both 
V-band photometry and spectral type.  Photometric errors are too small to see for most sources.}
\label{specvvmi}
\end{figure}

The X-ray active stars also follow clear sequences in an HR diagram (Figure \ref{specvvmi}, top panel) 
and in a diagram of V-I vs. spectral type (Figure \ref{specvvmi}, bottom panel).
Aside from two early-type stars, there is a clear relation between color and spectral 
type in the lower panel of Figure \ref{specvvmi}.  The X-ray active cluster 
sequence extends from V = 11, V-I= 0.4 to V = 23.5, 
V-I= 3.5.  Over this range, h Persei sources have spectral types 
betweeen B2 and M0.  
Most detected X-ray active stars 
are F5-G5 stars with V=16-20 and V-I = 1-1.5.  Early A stars with an intrinsic
V-I $\sim$ 0 have an observed V-I $\sim$ 0.7 consistent with 
the mean cluster reddening of E(B-V) $\approx$ 0.56 \citep{Bb88, Sl02}.

Using the same upper and lower bounds in V/V-I to identify probable cluster members  
lacking X-ray detections, Chandra detects less than 10\% of 
the h Persei cluster members.  Within 8.5' of the center of the Chandra/ACIS 
coverage, 1,621 stars with V=11-22 lie within 0.75 mags of the isochrone.
    Of these stars, 130 have Chandra detections.  
These stars thus have optical colors and magnitudes consistent 
with cluster membership.  The total fraction of X-ray detected cluster members with the 
same V, V-I limits and located within 8.5' of h Per is $\approx$ 
8\% (130/1621)\footnote{Given that the locus of giant stars crosses the cluster locus at 
V-I $\sim$ 1.7, we expect a small amount of contamination and thus some uncertainty in the 
true fraction of X-ray detected cluster members.}.  In comparison, 
there are 1,764 stars within 8.5' of h Per that lie off 
the isochrone; 28 of these have Chandra detections.  
Therefore, $\approx$ 1.6\% of nonmembers show evidence for X-ray activity (28/1764).

Converting from optical color and magnitude to stellar mass 
using the Geneva/Baraffe isochrone further constrains the fraction of cluster 
stars detected as a function of stellar mass.  Very few high mass stars (M=2 M$_{\odot}$--5 M$_{\odot}$, 
V/V-I = 0.52 --- 0.75) are detected (2.6\%, 6/230).  
A slightly larger fraction of subsolar-mass stars 
(0.6 M$_{\odot}$--1 M$_{\odot}$, V-I =1.87 --- 2.66) 
is detected (5.9\%, 24/407).  The detection rate for intermediate (1--2 M$_{\odot}$, 
V/V-I = 0.78 -- 1.87) mass stars is highest, $\approx$ 11\% (97/911).  

%
\subsection{X-ray Luminosity and Fractional X-ray Luminosity}
The vast majority of Chandra-detected sources with optical counterparts have 
optical magnitudes, colors, and spectra consistent with membership in h Persei.  We now 
analyze the X-ray properties of likely h Persei members.  X-ray active members 
are defined as those within 0.75 magnitudes of the 14 Myr isochrone\footnote{Technically, the 
upper limit should be larger than the lower limit because of binarity.  
For simplicity we make the upper and lower limits the same.  Using a more restrictive 
lower limit (e.g. 0.3 magnitudes) does not affect our results.}.  
We assume an extinction of E(B-V) $\sim$ 0.56, an age of 14 Myr, and 
a distance modulus of dM=11.85 for the cluster \citep{Sl02, Cu08b}. 

From these membership criteria, we investigate the X-ray luminosity function and
the distribution of fractional X-ray luminosities.  The X-ray luminosity function 
probes the absolute X-ray luminosity produced by the stellar chromosphere;  
the fractional X-ray luminosity function explores how X-ray luminous the star is 
compared to its photosphere.  We compare these luminosity functions 
with the stars' dereddened colors and spectral types to investigate how they 
correlate with stellar properties.   

The fractional luminosity, L$_{x}$/L$_{\star}$, is estimated from the observed L$_{x}$  
 and the bolometric luminosity, L$_{\star}$, derived from 
the well-constrained age, reddening, and distance for h Persei. 
Assuming that all h Persei stars are $\sim$ 14 Myr old and  
have E(B-V) $\sim$ 0.56, we evaluated the expected spectral type, mass, and bolometric
 luminosity from V,V-I.  The derived L$_{\star}$ for sources with V-I colors in between 
points on the Geneva/Baraffe isochrone grid is determined by interpolating 
between the points in theoretical log(L$_{\star}$) space.

\subsubsection{X-Ray and Fractional X-Ray Luminosity Functions}

\begin{figure}
\plotone{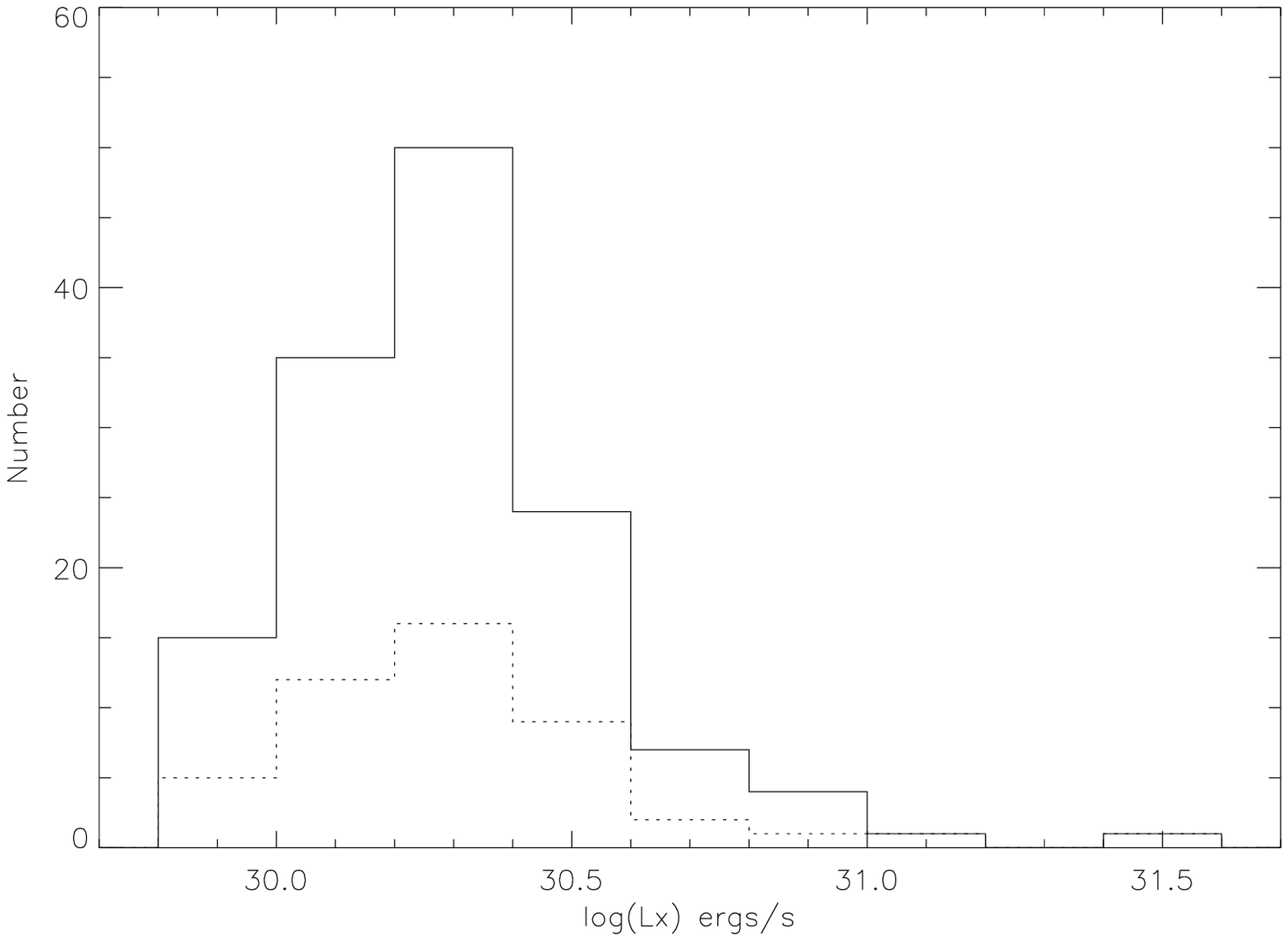}
\plotone{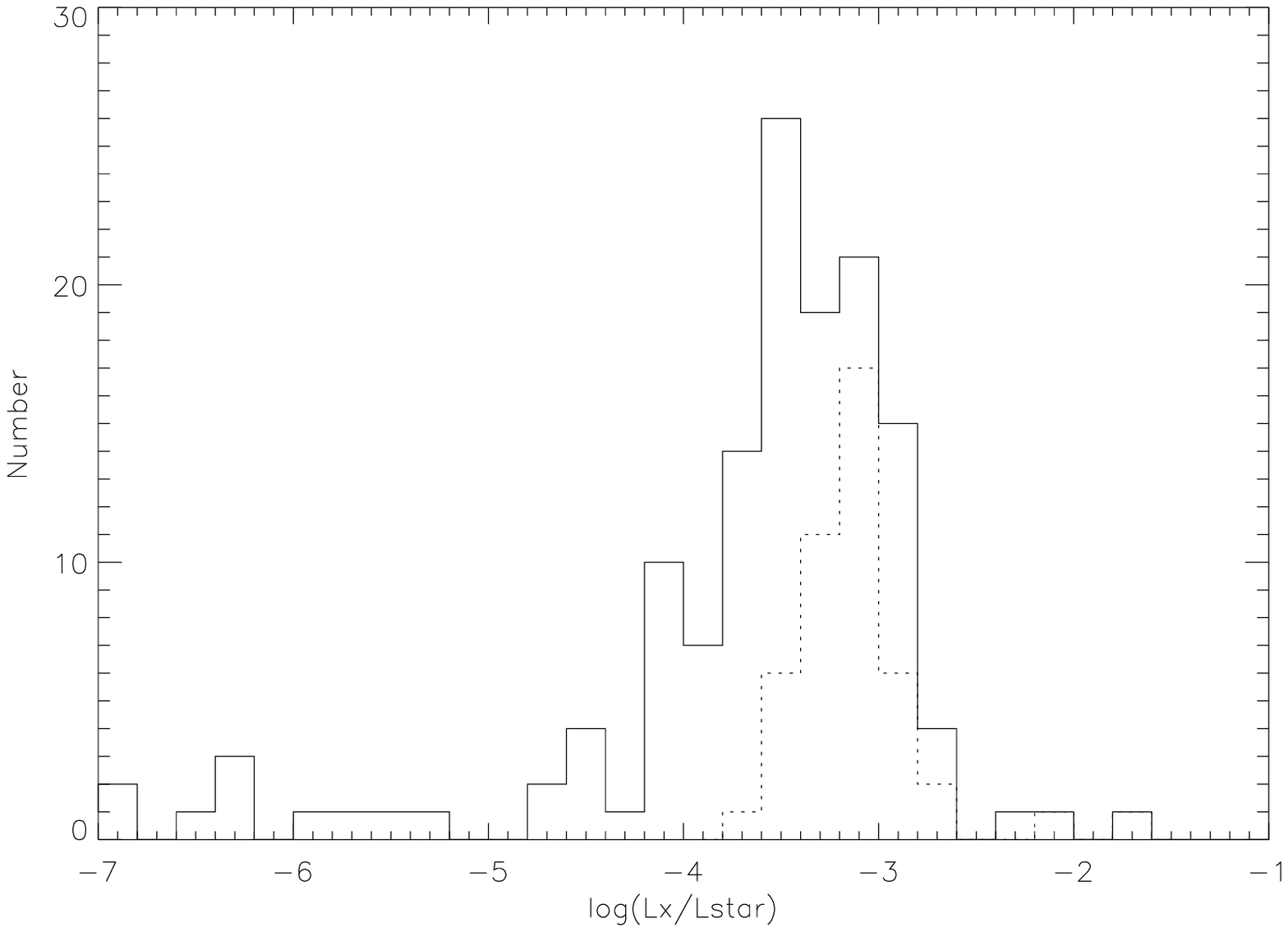}
\caption{Observed distributions of L$_{x}$ and L$_{x}$/L$_{\star}$ for the h Per stars with optical/IR counterparts.  
(top) Histogram of the L$_{x}$ for Chandra sources. 
  The number counts peak at log(L$_{x}$) = 30.3 ergs s$^{-1}$ for all stars (solid line) and 
at log(L$_{x}$) $\sim$ 30-30.4 ergs s$^{-1}$ for stars with M $\approx$ 1 M$_{\odot}$ (dashed lines).  
(bottom) Histogram of the fractional X-ray luminosity.  The fractional luminosity 
is peaked at log(L$_{x}$/L$_{\star}$) $\sim$ -3.5 for all stars (solid line) and at -3.2 for stars with 
M $\sim$ 1 M$_{\odot}$ (dashed line).  X-ray radiation contributes 
$\sim$ 10$^{-4}$--10$^{-3}$ of the total stellar flux for most h Per stars.  In both plots, Chandra sensitivity 
limits preclude us from fully sampling the entire X-ray luminosity function and fractional X-ray luminosity function.}  
\label{lxdist}
\end{figure}

Figure \ref{lxdist} shows the observed X-ray luminosity function for h Persei.  
The number counts peak at log(L$_{x}$, ergs s$^{-1}$) $\sim$ 30.3 
(2 $\times$ 10$^{30}$ ergs s$^{-1}$).  To compare the peak to predictions 
for the limiting X-ray luminosity, we follow \citet{Fei02, Fe05} who compute the 
limiting L$_{x}$ for a given exposure time, source distance, and n$_{H}$.  
For the full 0.5--8 keV band, log(L$_{x, lim}$, ergs s$^{-1}$)
 $\sim$ 30.3, in agreement with the observed peak.

We also show the 
distribution for the subset of stars with V-I=1.585-2.225 (dotted line).  
At 14 Myr, these stars should have M$_{\star}$ $\sim$ 0.9--1.2 M$_{\odot}$ and have 
G spectral types on the main sequence.  The distribution 
for 0.9--1.2 M$_{\odot}$ stars is similar to the function for 
the entire population, though it lacks any stars with log(L$_{x}$) $\ge$ 31.

Because Chandra detects $\lesssim$ 10\% of all cluster stars,
members without detections have lower X-ray luminosities.  X-ray luminosity 
functions for both young and older clusters have many stars with 
log(L$_{x}$ ergs s$^{-1}$) $<<$ 30.3 \citep[e.g.][]{Pr205, Je06}.  Thus, the 
distribution of L$_{x}$ shown in Figure \ref{lxdist}
identifies the completeness limit in L$_{x}$, and the true
X-ray luminosity function peaks at log(L$_{x}$ ergs s$^{-1}$) $<<$ 30.3.

The distribution of L$_{x}$/L$_{\star}$ is peaked 
at L$_{x}$/L$_{\star}$ $\sim$ 10$^{-3.5}$ (Figure \ref{lxdist}b).
Because the vast majority of cluster stars are undetected, the true fractional 
X-ray luminosity function is likely peaked towards lower values.
Three stars have a much higher fractional luminosity ($\approx$ 10$^{-2}$), and several have 
a lower fractional luminosity of $\approx$ 10$^{-3.8}$--10$^{-4.8}$.  

Five sources-- all early-type, high-mass stars-- have an 
extremely low fractional luminosity of $\approx$ 10$^{-6}$--10$^{-7}$.  
High-mass stars produce X-ray emission from wind shocks, not
a magnetic dynamo as in solar/subsolar-mass stars, which yields far lower
L$_{x}$/L$_{\star}$ \citep{Ste05}.  While an unseen low-mass companion 
could be responsible for the X-ray emission, this is an unlikely 
source for X-ray emission around high-mass stars in h Per.  B5 stars 
with L$_{x}$/L$_{\star}$ $\approx$ 10$^{-6}$ typically have V $\approx$ 12; 
G stars with L$_{x}$/L$_{\star}$ $\approx$ 10$^{-3.5}$ have V $\approx$ 18.  The bolometric corrections 
for B5 stars are BC $\approx$ -1.6; the corrections for G0 stars are BC $\approx$ -0.2.  Therefore, B5 stars are $\approx$ 
7.4 magnitudes brighter than G stars.  Thus, the fractional X-ray luminosity of a G0 star 
in a G0/B5 binary needed to yield an observed L$_{x}$/L$_{\star}$ for the system 
of $\approx$ 10$^{-6}$ is $\approx$ 10$^{-3}$.  Chandra detects only the most 
X-ray luminous G stars in h Per, which have L$_{x}$/L$_{\star}$ $\le$ 10$^{-3}$.  
Because $\gtrsim$ 90\% of cluster G stars are less X-ray luminous, the likelihood 
that an unseen companion generates the X-ray flux observed from B stars is low.

Deeper X-ray observations would yield a better measure of the X-ray luminosity 
function and fractional X-ray luminosity function 
in h Per.  In the younger Orion Nebula Cluster, \citet{Pr05} 
detect $\gtrsim$ 90\% of the 0.9-1.2 M$_{\odot}$ stars with typical 
L$_{x}$/L$_{\star}$ $\sim$ 10$^{-5}$ to 10$^{-2}$ and a median L$_{x}$/L$_{\star}$ 
$\approx$ 10$^{-3}$ to 10$^{-3.5}$.  At the distance of h Persei, many of these 
stars would lie below the Chandra detection limit for G-type stars.  
Thus, the true X-ray and fractional X-ray luminosity functions 
are likely peaked at lower L$_{x}$ and L$_{x}$/L$_{\star}$ values.  
\subsubsection{X-Ray Properties vs. Other Stellar Properties}
\begin{figure}
\plotone{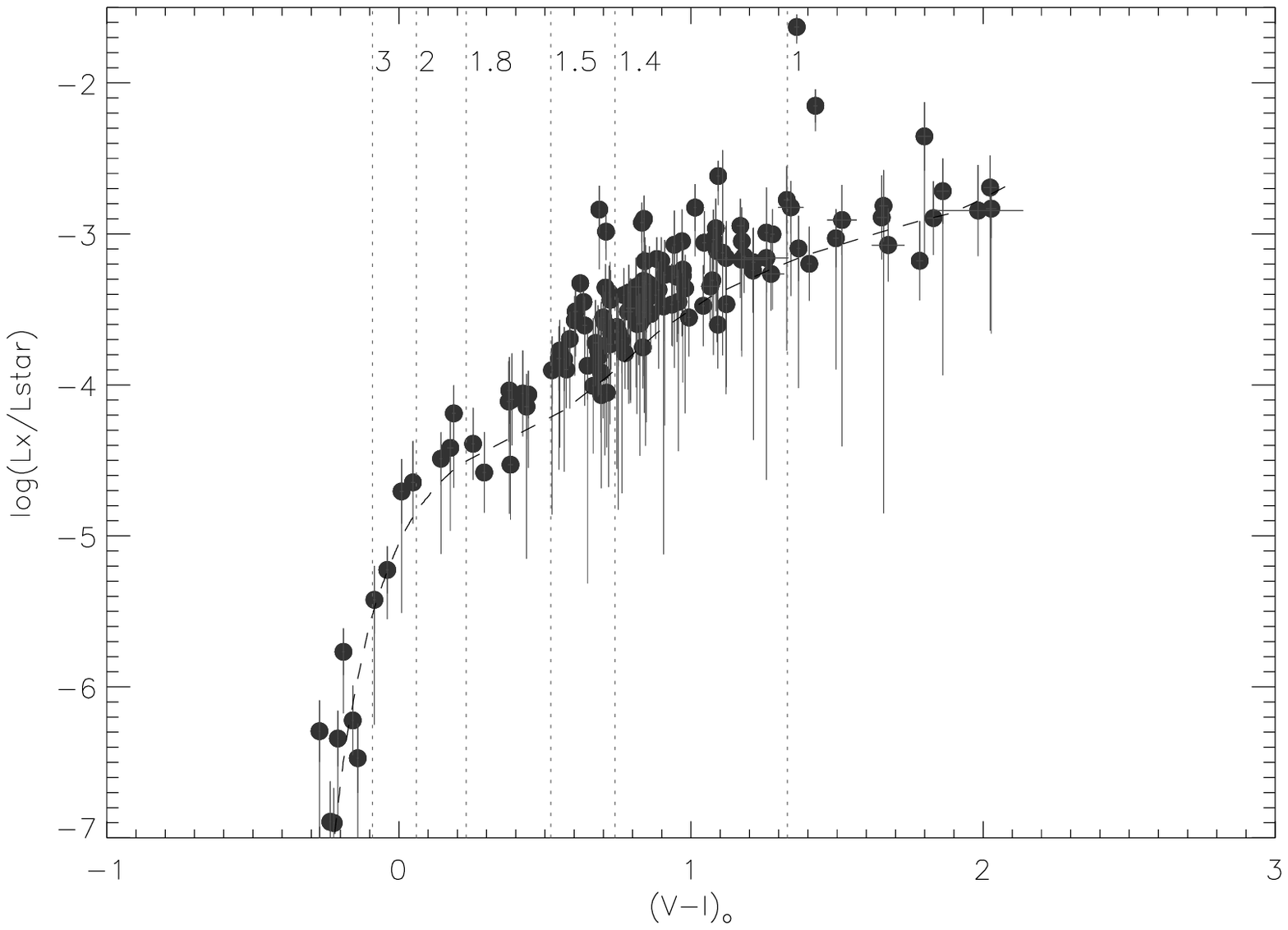}
\plotone{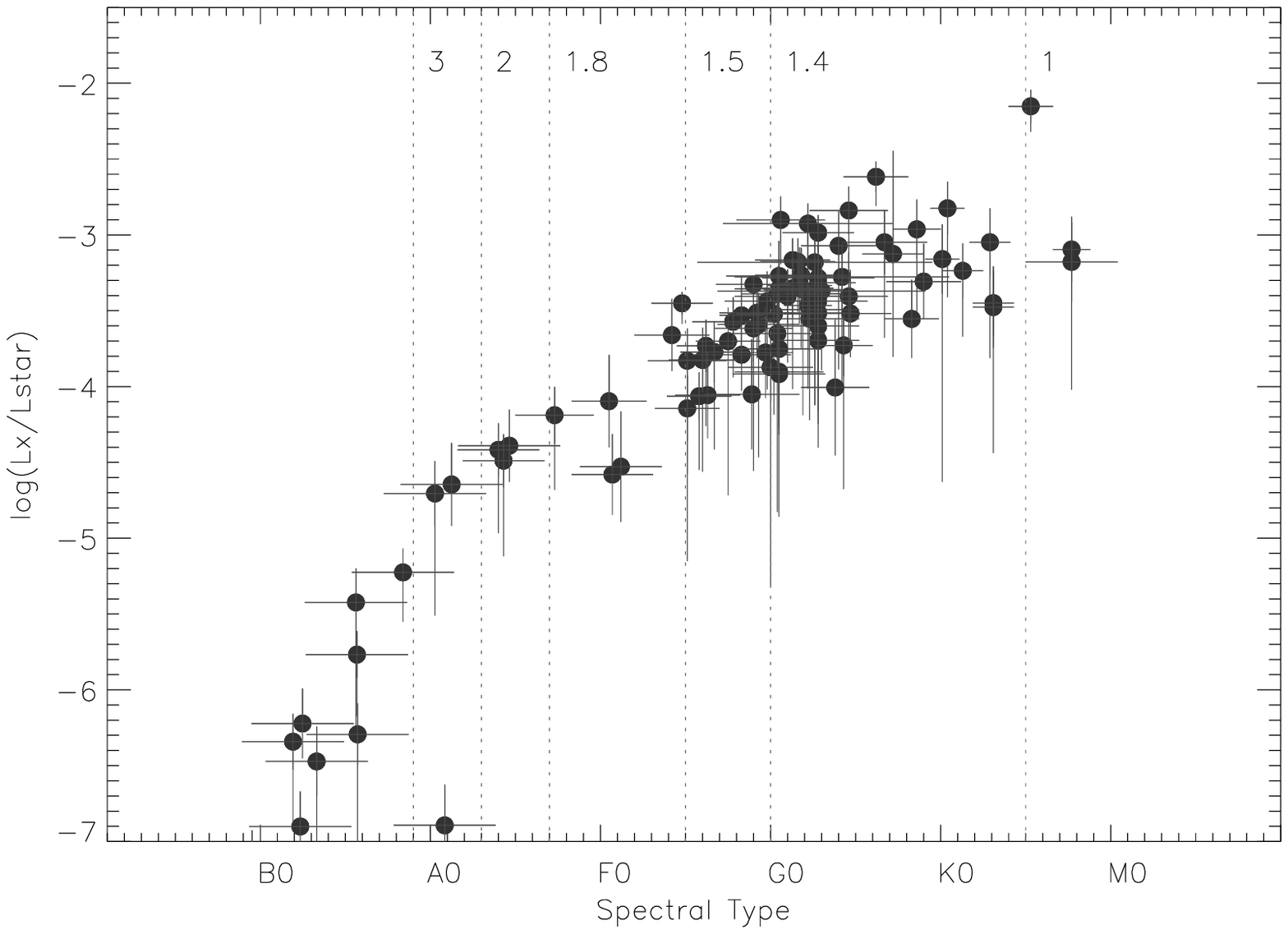}
\caption{Fractional X-ray luminosity as a function of dereddened V-I color (top panel) 
and spectral type (bottom panel).  The X-ray luminosity defines 
a clear sequence from bluer stars with a low luminosity to redder stars with a higher 
luminosity.  The nominal Chandra detection limit determined from \citet{Fe05} is shown as a dashed line.  
Uncertainties for luminosity and fractional luminosity shown in both panels 
are determined from the Sherpa fits; uncertainties for the color and spectral type are 
also shown.  Stellar masses for 14 Myr-old stars from \citet{Ba98} for low-mass stars and 
\citet{Si00} for high-mass stars are indicated by vertical dotted lines.
}
\label{xvsopt}
\end{figure}

The fractional X-ray luminosity correlates well with dereddened V-I color 
(Figure \ref{xvsopt}a).  Most sources occupy a region 
bounded by V-I = -0.2-1.3 and log(L$_{x}$/L$_{\star}$) $\sim$ 
-4 to -3.  Bright cluster stars with blue V-I colors 
have weak X-ray emission compared to their photospheres;  
  fainter stars with red V-I colors have relatively strong X-ray emission.  
Figures \ref{xvsopt}a and b show that the five extremely low fractional 
luminosity sources identified in Figure \ref{lxdist}b are mostly B0-B5 stars. 
The lack of low-luminosity late-type stars is likely due to 
the Chandra sensitivity limits, which are shown as a dashed line. 
However, high-luminosity, early-type stars (log ($L_{x}$/L$_{\star}$) $\sim$ -3) 
with V-I $\sim$ -0.2 -- 0.3 are absent in h Persei and should 
have been easily detected.

Because brighter clusters stars at a given age have earlier spectral types 
than fainter cluster stars, the fractional X-ray luminosity of h Persei stars also correlates with 
spectral type (Figure \ref{xvsopt}b).  Cluster stars with V-I $\sim$ -0.2 -- 0.3 have 
spectral types between $\sim$ B0 and F5.  Though h Persei contains 
thousands of stars with these spectral types, none of them 
have L$_{x}$/L$_{\star}$ $\gtrsim$ 10$^{-3.5}$.

 In summary, Figure \ref{xvsopt} shows that stars with small L$_{x}$/L$_{\star}$ 
($\approx$ 10$^{-7}$--10$^{-5}$) are early type (B/A), higher-mass stars 
while those with larger L$_{x}$/L$_{\star}$ ($\approx$ 10$^{-5}$--10$^{-3}$)
are intermediate type/mass stars (FGK).  The B and A stars with X-ray emission 
are more massive than $\approx$ 2 M$_{\odot}$ at 14 Myr \citep{Ba98, Si00}.  
The F--K stars with higher fractional luminosity have masses of $\approx$ 1--2 M$_{\odot}$. 
While Chandra likely detects only the most X-ray luminous cluster stars, these data 
provide some constraints on the X-ray luminosity function, specifically its upper 
envelope as a function of spectral type.

\subsection{X-Ray Properties and IR Excess from Warm Debris Disks}
The Double Cluster, h and $\chi$ Persei, contains the largest known 
population of warm debris disks \citep{Cu07a, Cu07b, Cu08a, Cu08b}.  
Ongoing terrestrial planet formation is the most likely source for 
the observed debris \citep{Kb04}.  Here we analyze 
the X-ray properties of Chandra-detected h Persei sources with warm debris disks.
We focus on a) whether X-ray luminosity correlates with the amount of debris emission 
and b) identifying the X-ray luminosity of X-ray bright stars that 
are likely forming terrestrial planets.

\begin{figure}
\plotone{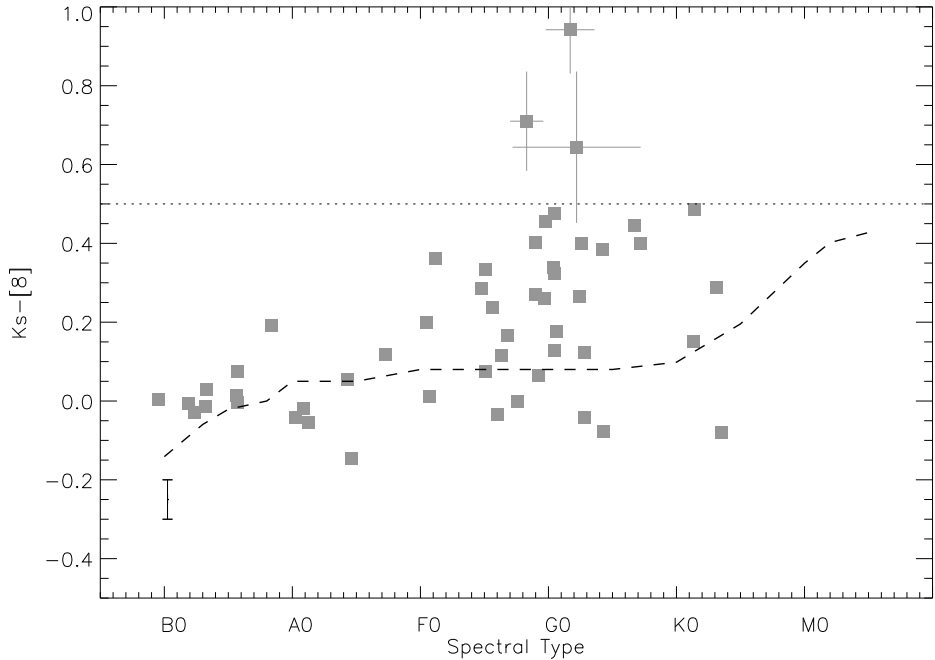}
\caption{K$_{s}$-[8] vs. spectral type for optically-detected Chandra sources consistent with cluster membership.   
The dashed line shows the locus of photospheric K$_{s}$-[8] colors 
from the SENS-PET tool (available on the Spitzer Science Center website) 
using the Kurucz-Lejeune stellar atmosphere models.  The dotted line shows the division between 
photospheric and excess sources adopted here.  Error bars for excess sources are shown explicitly; 
the error bar in the lower left corner indicates a typical 
uncertainty in the K$_{s}$-[8] color ($\sigma$ $\sim$ 0.1) for all sources. 
  Cooler stars have a larger range of K$_{s}$ -[8] colors.  At least 
three of the X-ray active F and G stars have red 2MASS-IRAC colors consistent with 
warm, terrestrial zone dust emission.  
}  
\label{stypekm4}
\end{figure}

Figure \ref{stypekm4} shows the spectral types of Chandra sources
 vs. their observed K$_{s}$-[8] colors.  X-ray bright stars with 
IRAC detections cover a wide range of spectral types (B0-K3).  
Most stars have spectral types earlier than G3 with K$_{s}$-[8] $\approx$ 0--0.3. 
To compare these colors with the predicted colors of stellar photospheres, 
we add the locus of stellar K$_{s}$-[8] colors for B0--M0 stars from 
the Kurucz-Lejeune stellar atmosphere models using the SENS-PET tool 
available on the Spitzer Science Center website\footnote{http://ssc.spitzer.caltech.edu/tools/senspet/}.  
The locus of photospheric 
colors is nearly independent of spectral type from A stars through G stars 
(K$_{s}$-[8] $\approx$ 0), covering almost the entire spectral type range of Chandra sources.

Figure \ref{stypekm4} also reveals four Chandra sources 
with IR excess emission from circumstellar dust.
  \citet{Cu07a} identified 8 $\mu m$ excess sources 
as those with K$_{s}$-[8] $\ge$ 0.4 + $\sigma$[8], where 
$\sigma$[8] is the typical uncertainty in the 8 $\mu m$ 
magnitude.  For FGK stars with typical [8] $\sim$ 13.75,  
$\sigma$[8] $\sim$ 0.1 \citep{Cu07a}.  Three stars have K$_{s}$-[8] 
colors redder than $\sim$ 0.5 and thus have 8 $\mu m$ 
excess consistent with warm, terrestrial zone dust.
A fourth star without a spectral type also has 
K$_{s}$-[8] $>$ 0.5.  Thus, at least 4 Chandra detections 
have excess emission at 8 $\mu m$.

The frequency of 8 $\mu m$ excess emission among 
Chandra detections is comparable to the fraction in the entire 
cluster.  The frequency of Chandra-detected stars  
 with 8 $\mu m$ excess is 4/123 (3.3\% $\pm$ 1.7\%.  The frequency for 
Chandra-detected cluster members is 4/106 (3.7\% $\pm$ 1.9\%).  
For all F0-G5 stars in h and $\chi$ Persei, the fraction ranges 
from $\approx$ 4\% to 8\% \citep{Cu07a, Cu08b, Cu08c}.
Given the small sample of X-ray bright stars with evidence for warm dust, 
the similarity in excess frequency between them and stars lacking Chandra 
detections should be considered a tentative result.

\begin{figure}
\plotone{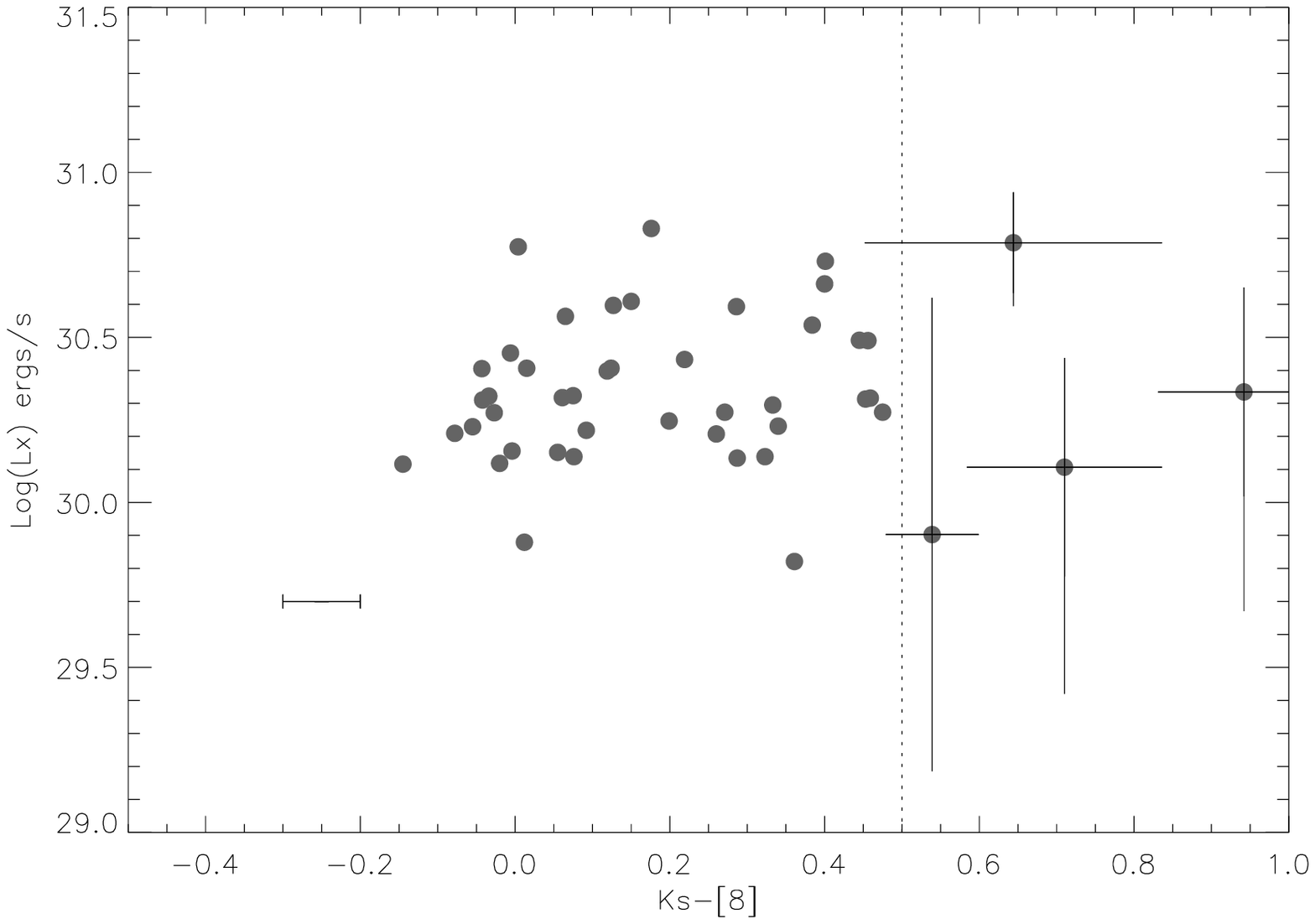}
\plotone{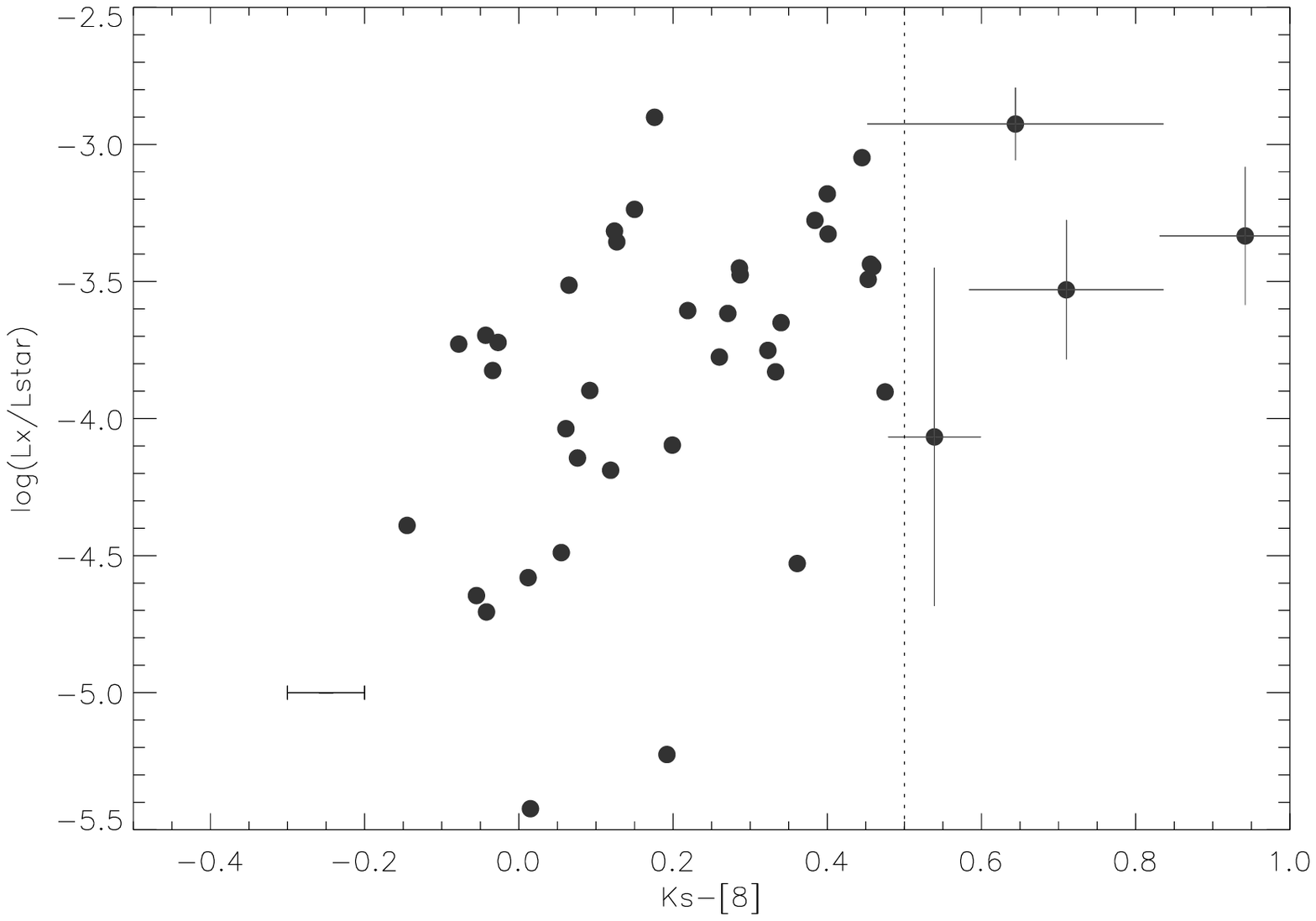}
\caption{The relationship between X-ray luminosities and IR excess emission.  
L$_{x}$ vs. K$_{s}$-[8] (top panel) and log(L$_{x}$/L$_{\star}$) vs. K$_{s}$-[8] (bottom panel) 
for sources detected by Chandra, 2MASS, and the 8$\mu m$ filter in IRAC.  
The dotted line 
corresponds to the division between excess emission and no excess emission.  
The uncertainties in (fractional) luminosity and photometry are shown for excess sources; the 
typical K$_{s}$-[8] uncertainty is also shown (lower left-hand corner).
Most sources 
lie in a distribution between K$_{s}$-[8] =0 and 0.4.  About 3-4 stars may have 
IRAC excess emission indicative of warm, terrestrial zone dust. } 
\label{lxvwarm}
\end{figure}

Figure \ref{lxvwarm} compares the K$_{s}$-[8] colors and X-ray luminosities 
for Chandra-detected cluster members.  Most stars have log(L$_{x}$, ergs s$^{-1}$) $\sim$ 30-30.8.  
Sources with red K$_{s}$-[8] colors do not identify a unique space in this plot.  
Three 8 $\mu m$-excess stars have log(L$_{x}$, ergs s$^{-1}$) $\sim$ 30-30.4; a fourth has 
log(L$_{x}$, ergs s$^{-1}$) $\sim$ 30.8.   
The Spearman's rank test, a non-parametric measure of correlation, 
reveals that the distribution of L$_{x}$ with K$_{s}$-[8] color 
is consistent with a random distribution (probability $\approx$ 23\%, d=0.17).

Figure \ref{lxvwarm}b shows the distribution of fractional X-ray luminosity with 
K$_{s}$-[8] color.  Most IRAC-detected stars have L$_{x}$/L$_{\star}$ $\sim$ 
10$^{-4.75}$--10$^{-2.75}$, which are typical of the X-ray bright population 
as a whole.  Sources with 8 $\mu m$ excess define a slightly more 
narrow distribution between L$_{x}$/L$_{\star}$ $\sim$ 10$^{-4}$--10$^{-2.9}$.  
According to the Spearman's rank test, the probability that L$_{x}$/L$_{\star}$ 
is not correlated with the K$_{s}$-[8] color is $\approx$ 0.89\% (d=0.44).
Thus, there is a 2--3$\sigma$ correlation between X-ray flux and K$_{s}$-[8] color.

If this trend is real, the fractional X-ray 
luminosity probably does not affect the amount of debris emission.
\citet{Cu07a} and \citet{Cu08c} 
find that intermediate-type (FG) stars more frequently have 8 $\mu m$ 
excess than earlier stars.  Intermediate-type stars also have  
higher fractional X-ray luminosities regardless of their circumstellar 
environment (Figure \ref{xvsopt}b).  Therefore, the correlation between 
L$_{x}$/L$_{\star}$ and IR excess in Figure \ref{lxvwarm}b is likely 
due to the intrinsically larger fractional X-ray luminosities and higher
 frequencies of warm dust for intermediate-type stars compared to 
early-type stars. 

In spite of these results, we caution that they apply only to the \textit{X-ray bright} population 
of stars with warm dust, not the population of disk-bearing sources as a whole.  
Other stars with warm dust located within the Chandra coverage must be more 
X-ray faint.  Deeper Chandra data are required to more definitively probe the relationship between
 X-ray activity and IR excess.
\subsection{X-ray Active Stars and Planet Formation in h Persei}
If the 8 $\mu m$ excess emission is due to active terrestrial planet formation, 
we can estimate the X-ray flux incident on forming planets given 
the temperature and location of the dust.  
\citet{Cu08a} derive dust temperatures of $\approx$ 250-400 K for stars 
with 8 $\mu m$ excess.  Using the bolometric luminosity derived from the 
Geneva/Baraffe isochrone, we calculate the location of the dust assuming 
that dust grains are in radiative equilibrium with a temperature of $\approx$ 300 K.
Derived dust locations range from $\approx$ 0.92 AU to 1.36 AU.  
From the location of the dust, we calculate the X-ray flux.

The incident X-ray flux on objects at the dust locations
 ranges from 1.53 $\times$ 10$^{2}$ ergs cm$^{-2}$ s$^{-1}$
to 2.20 $\times$ 10$^{3}$ ergs cm$^{-2}$ s$^{-1}$.  These levels are equal to or larger than 
 the X-ray flux intercepted by the atmosphere of the hot Jupiter HD 209458b ($\sim$ 
200 ergs cm$^{-2}$ s$^{-1}$ for L$_{x}$ $\approx$ 1.1$\times$ 10$^{27}$ ergs s$^{-1}$ at 
0.045 AU, see \citealt{Pe08}).   Thus, a planet with a mass and radius similar to HD 209458b
located at $\approx$ 1 AU from an X-ray luminous G star in h Per may experience similar 
rates of atmospheric erosion of up to $\approx$ 10$^{10}$ g s$^{-1}$ \citep{Vid03, Vid04}.  

However, this level of atmospheric escape will not likely be sustained by 
any jovian planets in the terrestrial zones of h Per stars.  \citet{Pe08} indicate that the timescale for 
erosion of a significant fraction of a planet's mass is on the order of 100 Myr--1 Gyr.  
Over this timescale, the X-ray luminosity of the star will drop by several orders 
of magnitude \citep{Pr05}, which greatly reduces the rate of atmospheric escape.  

\section{Comparison With Other 1--100 Myr-old Chandra-observed Clusters}

X-ray emission from young stars is most often associated with a magnetic dynamo, the 
"$\alpha$-$\Omega$" dynamo \citep{Pa55, Ba61, Ba03}.  Differential rotation between 
an inner radiative core and an outer convective envelope drives the dynamo and powers 
chromospheric and coronal activity. For the \citet{Si00} isochrones of a 14 Myr-old
cluster, we expect stars with outer convective envelopes to have spectral types later
than $\sim$ G0 and stellar masses less than $\sim$ 1.4--1.5 $M_{\odot}$.

Early-type stars without convective envelopes do not produce a dynamo. Thus, these 
stars do not display stellar activity.  However, many early-type stars have large
X-ray luminosities with typical $L_x / L_{\star} \approx$ $4-400 \times 10^{-6}$ 
\citep[e.g.,][]{Chl89,Ste05}. Shocks in high velocity stellar winds are the most 
likely source of X-ray emission in these stars \citep[e.g.,][]{Feld97}. For h Per, 
we expect wind emission to dominate chromospheric activity for stars with spectral
types earlier than F0 and masses larger than $\sim$ 1.7 $M_{\odot}$.

Our results suggest that h Per contains both types of X-ray sources. Late-type stars
likely powered by the $\alpha$-$\Omega$ dynamo have $L_x / L_{\star} \lesssim$ 
$10^{-3}$ (Figure \ref{xvsopt}).  Stars with spectral types earlier than $\sim$ F0 have 
fractional X-ray luminosities comparable to typical wind-driven sources. These
fluxes are probably not produced by an unseen late-type companion (\S3.2.1). Thus,
shocks probably produce X-ray emission in these stars. 

In this section, we compare the X-ray properties of h Per stars with X-ray active 
stars in younger and older clusters studied by Chandra and XMM-Newton.  We divide 
our sample into "early" (B0-A3) and "intermediate" (FGK) type stars.  Comparing h Per 
data with results for other clusters then provides constraints on the evolution of 
X-ray activity from intermediate-type stars and stellar wind-driven emission from early-type stars.

\subsection{X-Ray Emission from Early-Type, High-Mass Stars (B0-A3)}
\citet{Ste05} 
divide early type X-ray active stars in the ONC into two groups. "Strong wind" stars have 
spectral types between O and B3 (M $\gtrsim$ 7 M$_{\odot}$); "weak wind" stars have spectral types between 
B5 and A9 (M $\sim$ 2--7 M$_{\odot}$).  More than half (16) ONC early-type stars show clear evidence for X-ray emission.
 Orion stars with spectral types earlier than B3 are strong-wind stars with X-ray emission.
About 64\% of stars with spectral types between B5 and A9 are detected by Chandra.

At 14 Myr, assuming that the division between strong and weak-wind stars is set by stellar mass,
 strong wind sources in h Per have spectral types earlier than B3.  Weak wind 
stars have spectral types between B3 and A3.  Chandra detects 9 h Per stars earlier than A3.
None of these stars correspond to Be stars or candidate Be 
stars  identified by \citet{Bk05}, \citet{Sl02}, and \citet{Cu08a} and none show H$_{\alpha}$ emission 
 characteristic of Be stars.  Compared to the total population of B0-A3 stars 
within 8.5' of h Persei (where V-I $\le$ 0.76), $\approx$ 4\% are detected by Chandra.  
About 5.7\% (6/106) of stars earlier than B3 (V-I $\lesssim$ 0.4) are detected, while 2.8\% 
(6/217) of stars between B3 and A3 are detected.  Other early-type cluster members 
either lack X-ray activity or have wind-driven X-ray luminosities less than 
 $\approx$ 10$^{30}$ ergs s$^{-1}$.

If the X-ray luminosity function for strong and weak-wind sources does not evolve 
from $\sim$ 1 Myr to $\sim$ 14 Myr, Chandra should have detected more 
weak wind and strong-wind h Per stars.  Most (6/9) of the strong-wind stars in Orion 
have X-ray luminosities greater than the peak in L$_{x}$ for h Per ($\sim$ 30.3 ergs s$^{-1}$).  
Similarly, 44\% (3/7) of the weak-wind Orion stars have L$_{x}$ $\ge$ 30.3 ergs s$^{-1}$.
Although sensitivity limits for h Per 
observations preclude more detailed comparisons between the ONC and h Per
luminosities for early-type stars, these results may suggest that the X-ray luminosities 
for early-type, high-mass stars at 14 Myr are weaker than they are at 1--2 Myr.  
\subsection{X-Ray Emission from $\lesssim$ 3 M$_{\odot}$ (FGK) Stars}
Recent Chandra and XMM-Newton observations of young stellar clusters 
 provide some constraints on the evolution of X-ray activity 
as a function of age and stellar mass for intermediate and low-mass 
stars.  Independent of stellar mass, the typical fractional X-ray luminosity 
of 1 Myr-old stars in the ONC is L$_{x}$/L$_{\star}$ $\approx$ 10$^{-3}$ \citep{Pr205}.  For most 
young stars, the stellar rotational velocity increases with stellar mass, so  
 L$_{x}$/L$_{\star}$ is independent of rotation \citep{Pr05}.  By 
analogy with dynamo models for X-ray activity, the fractional X-ray luminosity 
for these stars is "saturated".

By $\approx$ 40-50 Myr, when intermediate-mass stars are on the main sequence, 
the stellar rotation rates of intermediate-mass
 have declined significantly \citep{Pr96, Je06} and correlate better with X-ray flux.  
Stars in 50 Myr-old $\alpha$ Persei \citep[][Fig. 5]{Pr96} show 
a systematic decline in L$_{x}$/L$_{\star}$
from 1.2 M$_{\odot}$ (B-V $\sim$ 0.6, G1 at 50 Myr)
to 1.4 M$_{\odot}$ (B-V $\sim$ 0.4, F5 at 50 Myr).
Cluster stars with masses $<$ 1.2 M$_{\odot}$ have 
L$_{x}$/L$_{\star}$ $\sim$ 10$^{-3}$ and thus have saturated 
X-ray emission.  The distribution of L$_{x}$/L$_{\star}$ for 
NGC 2547 (38 Myr) shows a similar trend \citep{Je06}.

At ages $\gtrsim$ 100 Myr, subsolar-mass stars 
leave X-ray saturation.
A wide range of stars in the Pleiades (M $\le$ 0.85 M$_{\odot}$;
K3 at 100 Myr) are intrinsically much fainter in X-rays than 
younger stars with the same mass. 
Finally, older field stars more massive than $\gtrsim$ 0.1 M$_{\odot}$ also
exhibit a decline in X-ray luminosity relative to younger clusters \citep{Pr205,Pr05}. 
L$_{x}$/L$_{\star}$ ranges from 
10$^{-4.5}$ to 10$^{-8}$ for solar-mass stars and 10$^{-4}$ to 10$^{-6.5}$ for 
subsolar-mass stars.  The evolution of 
fractional X-ray luminosity vs. time shows a systematic decline and 
is consistent with the \citeauthor{Sk72} braking track \citep[L$_{x}$(t) $\propto$ t$^{-1/2}$]{Sk72}.

At least two mechanisms are responsible for the evolution of 
L$_{x}$/L$_{\star}$ as a star contracts onto the main 
sequence.  Rotational spin-down of a star 
reduces differential 
rotation at the radiative zone/convective zone boundary.  
Because differential rotation is responsible for 
driving the magnetic dynamo that produces X-ray emission, 
X-ray activity diminishes \citep{Pr05}. 

Intermediate and high-mass stars also 
undergo significant structural changes 
as they evolve onto the main sequence.  
At the age of 
the Orion Nebula Cluster ($\sim$ 2 Myr), main sequence 
A stars (1.7--2.5 M$_{\odot}$) have spectral types between K1 and K4 \citep{Si00}.
From 2 Myr to 30 Myr, $\sim$ 2 M$_{\odot}$ stars evolve in spectral type from 
K3 to G8 (5 Myr) to A3 (30 Myr).  Main sequence F3 stars ($\sim$ 1.5 M$_{\odot}$) also 
undergo substantial changes in their internal structure, evolving from 
at $\sim$ K5 at 2 Myr, to K0 at 10 Myr, to F3 at 30 Myr.
As stars evolve from G/K stars to A/F stars, their outer convective zones 
shrink and then disappear.  The disappearance of their convective zones 
eliminates magnetic dynamo-induced X-ray activity \citep[see ][]{Gi86, Ste05, Pr05}.

The X-ray detections for stars in h Per place useful constraints
on both of these mechanisms. For h Per stars with masses $\sim$
1.0--1.5 $M_{\odot}$, the maximum in log $L_x/L_{\star}$ ($\sim$
$-$2.8 to $-$3.0) is similar to the maximum observed in the ONC.
Thus, there is no clear evidence for evolution in X-ray activity among
1--1.5 $M_{\odot}$ stars from $\approx$ 1~Myr to $\approx$ 14 Myr.
Deeper Chandra data are required to probe the X-ray luminosity 
function of solar and subsolar-mass stars at smaller 
L$_{x}$/L$_{\star}$ and to constrain the evolution 
of their X-ray activity.

For more massive stars, there is clear evidence of evolution in
X-ray activity. At $\approx$ 14 Myr, h Per stars with A0 to $\sim$ G0
spectral types have masses of 3 $M_{\odot}$ to 1.5 $M_{\odot}$.
In the ONC, stars with these masses have a maximum log $L_x/L_{\star}$
$\approx$ $-$2.7 to $-$3.0. This maximum is independent of spectral
type. In h Per, the maximum $L_x/L_{\star}$ clearly declines with
spectral type. Our Chandra data yield an approximate relation log
$(L_x/L_{\star})_{max}$ $\approx$ $-3.5 - (M_{\star} / 1.5 M_{\odot})$
for 1.5--3 $M_{\odot}$ stars. Thus, the fractional X-ray luminosity
declines by a factor of roughly 30 (3) for 3 (1.5) $M_{\odot}$ stars.

 Pre-main sequence stellar evolution is the simplest explanation
for the evolution of log $(L_x/L_{\star})_{max}$ in 1.5-3 $M_{\odot}$
stars. As these stars evolve to the main sequence, their
effective temperatures increase and they lose their convective
envelopes. Stars with masses $\gtrsim$ 1.8 $M_{\odot}$ have
spectral types earlier than A7 and lack the convective atmospheres
need for magnetic dynamo-driven X-ray activity. Thus, the X-ray
activity of these stars may be similar to the weak wind sources at
earlier spectral types.  Stars with masses $\sim$ 1.5--1.8 $M_{\odot}$
have $\sim$ A7--G0 spectral types and much smaller convective atmospheres
than G-type stars. These stars can drive weak dynamos and thus have
larger fractional X-ray fluxes than earlier type stars. In both sets
of stars, the decline in X-ray activity appears linked to changes in
stellar structure as the stars evolve from K-type stars at $\sim$
1--2 Myr to A-type stars on the main sequence.

 Deeper Chandra data in h and $\chi$ Per and other 5--50 Myr-old clusters
can test these conclusions. If stellar evolution is responsible for the
decay of X-ray emission among more massive stars approaching the main
sequence, this decay should be correlated with spectral type and
uncorrelated with stellar rotation. Chandra observations of other
young clusters are needed to see how $(L_x/L_{\star})_{max}$ and the
X-ray luminosity function of massive stars evolves with cluster age. If
changes in the magnetic dynamo are responsible for the decay of X-ray
activity among lower mass stars, deeper observations of (i) 10--20 Myr
clusters (to measure the X-ray luminosity function for 0.1--1.5 $M_{\odot})$
stars) and (ii) 20--50 Myr clusters (to measure evolution in
$(L_x/L_{\star})_{max}$) are required.

\section{Summary}
We analyze the first Chandra survey of 
h Persei, part of the Double Cluster in Perseus.  By matching 
Chandra-detected sources with optical/IR photometry and spectra, 
we identify X-ray active cluster members.  Using the well 
constrained age and distance to h Per, we compute the bolometric 
luminosity of h Per stars and compare this luminosity to the 
derived X-ray luminosity.  Comparing the X-ray luminosity and fractional 
X-ray luminosity to optical colors, spectra, and 
near-to-mid IR colors, we investigate the connection between X-ray 
activity and stellar/circumstellar properties.

These analyses yield the following results:
\begin{itemize}
\item There is a clear correlation of V, V-I with spectral 
type for the X-ray detections.  
The position of X-ray active stars relative to the observable 
cluster locus in V, V-I and their locus in V vs. spectral type 
shows that the vast majority of these stars are likely cluster members. 

\item Within the ACIS coverage, Chandra detects 8\% of cluster 
stars more massive than $\approx$ 0.6 M$_{\odot}$.  About 10\% of 
1--2 M$_{\odot}$ stars are detected while $\sim$ 3\% of 2--5 M$_{\odot}$ stars 
and $\approx$ 6\% of 0.6--1 M$_{\odot}$ stars are detected.  Thus, 
the sources detected by Chandra likely correspond to the 
bright end of a much larger population of X-ray active stars.

\item The observed distribution of X-ray luminosities for Chandra-detected h Per 
sources peaks at log(L$_{x}$) $\sim$ 30.3.  
Because of the sensitivity limitations 
of Chandra, the true X-ray luminosity function likely peaks at a lower 
luminosity. 

\item The observed distribution of fractional X-ray luminosities correlates 
well with the V-I color and with spectral type.  Early-type 
stars (M $\gtrsim$ 2 M$_{\odot}$) have a very small fractional luminosity (log(L$_{x}$/L$_{\star}$) $<$ 
-4 to -5); L$_{x}$/L$_{\star}$ is much larger for G and K stars (M $\sim$ 1--1.4 M$_{\odot}$,
(log(L$_{x}$/L$_{\star}$) $\sim$ -4 to -3).  

\item There is no evidence for an X-ray saturated population of 
stars more massive than $\sim$ 1.5 M$_{\odot}$.  In the Orion 
Nebula Cluster, the most X-ray active 
1.5--3 M$_{\odot}$ stars have log(L$_{x}$/L$_{\star}$) $\approx$ -3 
independent of stellar mass.  In h Persei, the most X-ray active 1.5--3 M$_{\odot}$ 
stars have log(L$_{x}$/L$_{\star}$) $\approx$ -3.5 - M$_{\star}$/1.5M$_{\odot}$. 
  Changes in stellar structure as these 
stars contract onto the main sequence likely play a critical role 
in this decline of X-ray activity.

\item From our sample of X-ray bright stars, the 
presence of warm, terrestrial zone circumstellar dust is not correlated with 
X-ray luminosity.  While stars with higher fractional X-ray luminosity 
are more likely to have warm dust, this trend is likely due to the higher 
frequency of warm dust around intermediate-type stars than around early-type stars.  

\item The warm circumstellar dust emission from X-ray active stars studied in 
this paper likely results from debris emission due to terrestrial planet formation 
\citep[e.g.][]{Cu07a, Cu07b, Cu08a, Cu08c}.  By computing the X-ray luminosity 
of these sources and assuming a dust temperature of $\approx$ 300 K, we 
derive an incident X-ray flux on forming terrestrial planets in systems 
with X-ray bright stars.  This flux is very small compared to stellar irradiation and the 
radiated flux from terrestrial zone dust.  Thus, X-ray emission
 is unlikely to affect the detectability of warm debris disks.  

\end{itemize}
These data show that h Persei is important in understanding the 
evolution of X-ray activity for a range of stellar masses and provides 
connections between X-ray activity and terrestrial planet formation.  
Deeper Chandra observations of h Persei can better constrain 
the X-ray luminosity function to lower stellar masses, provide a 
much larger sample from which to compare X-ray activity and 
planet formation, and more clearly reveal the characteristic 
X-ray environment experienced by forming terrestrial planets.
Deeper Chandra observations will also yield enough counts for 
X-ray bright sources to study the evolution of coronal temperatures 
in intermediate-mass stars.

\acknowledgements  We thank Leslie Hebb and Simon Hodgkin for KPNO observing and 
Nelson Caldwell, Susan Tokarz, Perry Berlind, and Mike Calkins 
for scheduling, taking, and reducing the Hectospec spectra.  
We also thank Tom Barnes and Rebecca Cover for early work on this project.  
Comments from the referee, Leisa Townsley, greatly improved the manuscript.
This work is supported by 
 NASA Astrophysics Theory grant NAG5-13278, Spitzer GO grant 1320379, and 
NASA TPF Foundation Science grant NNG06GH25G.  T.C. received support from a SAO Predoctoral Fellowship.
N.R.E., B.D.S, and S.J.W. are supported by NASA contract numbers NAS8-03060 and G0-6007A.
{}
\begin{deluxetable}{lllllllllllllllllll}
 \tiny
\rotate
\setlength{\tabcolsep}{0.015in}
\tabletypesize{\tiny}
\tablecolumns{14}
\tablecaption{Chandra/ACIS Detections}
\tiny
\tablehead{{ID}&{RA}&{DEC}&{Raw Counts}&{Net Counts}&
{HR1}&{HR2}&{HR3}&{n$_{H}$}&
{kT}&{$\sigma$(kT)}&{Flux$_{unabs.}$}&{Flux$_{abs.}$}&{Red.($\chi^{2}$)}\\ &
&{}&{}&{}&{}&{}&{}&{10$^{22}$ cm$^{-2}$}&{(keV)}&{(keV)}&{ergs s$^{-1}$ cm$^{-2}$}&{ergs s$^{-1}$ cm$^{-2}$}&{}}
\startdata
1& 34.9295& 57.1005& 34& 31.7& -0.1593& -0.2562& 0.0054&0.3&3.903& -999& 5.49E-15& 3.81E-15& 0.572\\
2& 34.9195& 57.0918& 6& 3.2&  -999 &-999& -999& 0.3&0.216& 0.044& 9.38E-15& 1.26E-15& 0.302\\
3& 34.9152& 57.0597& 28& 20.3& -999& -0.7477& -999&0.3& 0.913& 0.429& 4.25E-15& 1.84E-15& 0.593\\
4& 34.9039& 57.0547& 26& 20& -0.5233& -0.0469& -0.6703&0.3& 1.54& 0.641& 4.01E-15& 2.15E-15& 0.487\\
5& 34.8843& 57.1138& 17& 15.2& -0.7636 &-0.6217 &-0.4442&0.3& 1.599& 0.14& 2.54E-15& 1.38E-15& 0.532\\
\enddata
\tablecomments{Sources detected by Chandra/ACIS.  The hydrogen column, n$_{H}$, is fixed at 3$\times$ 10$^{21}$ cm$^{-2}$ 
based on the well-constrained reddening to h Persei.  The temperature (kT), normalization, and x-ray flux (Flux$_{unabs.}$) 
are determined by Sherpa.}
\label{Chandradet}
\end{deluxetable}

\begin{deluxetable}{lllllllllllllllllll}
 \tiny
\rotate
\setlength{\tabcolsep}{0.02in}
\tabletypesize{\tiny}
\tablecolumns{19}
\tablecaption{Catalog of Chandra sources with Optical Counterparts}
\tiny
\tablehead{{ID}&{RA}&{DEC}&{Net Counts}&{$\sigma$(Net Counts)}&{log(L$_{x}$) ergs s$^{-1}$}&{ST}&{$\sigma$(ST)}&{J}&{H}&{K$_{s}$}&{[3.6]}&{[4.5]}&
{[5.8]}&{[8]}&{V}&{$\sigma$(V)}&{I}&{$\sigma$(I)}}
\startdata
3& 34.9151&57.0597&20.3&5.83&30.47& 99&99&14.026&13.536&13.340&13.144&13.121&13.151&13.425&16.512& 0.001&15.050& 0.001\\
4& 34.9035&57.0546&20.0&5.74&30.45&48.6&1.4&15.883&15.235&14.932&14.870&99&99&99&19.094& 0.004&17.309& 0.003\\
5& 34.8843&57.1137&15.2&4.36&30.25&30.5&2.2&13.897&13.521&13.473&13.246&99&13.425&13.274&15.872& 0.001&14.785& 0.001\\
7& 34.8709&57.1545&10.3&2.96&30.17&13.2&2&11.044&10.975&10.970&10.920&10.924&11.085&10.984&11.814& 0.001&11.174& 0.001\\
8& 34.8705&57.1777&17.2&4.94&30.39&99&99&15.624&14.951&14.595&14.482&14.462&99&99&18.771& 0.003&16.982& 0.003\\ 
\enddata
\tablecomments{Chandra sources with optical counterparts.  The spectral types (ST) correspond to numerical spectral types: 
10=B0, 20=A0, 30=F0, 40=G0, and 50=K0, etc.  The uncertainties in spectral types ($\sigma$(ST)) are given in subclasses.  
Sources without spectral types and/or photometry in a given filter have entries denoted with '99'.}
\label{Chandralist}
\end{deluxetable}

\epsscale{0.8}
\end{document}